\documentclass[fleqn,usenatbib]{mnras}

\usepackage{newtxtext,newtxmath}

\usepackage[T1]{fontenc}

\DeclareRobustCommand{\VAN}[3]{#2}
\let\VANthebibliography\thebibliography
\def\thebibliography{\DeclareRobustCommand{\VAN}[3]{##3}\VANthebibliography}

\DeclareMathOperator\erf{erf}
\DeclareMathOperator\erfc{erfc}

\usepackage{graphicx}	

\usepackage[dvipsnames]{xcolor}

\newcommand{\gaia}{{\it Gaia}}

\title[Starspots in the Pleiades and M67]{Starspots and Magnetism: Testing the Activity Paradigm in the Pleiades and M67}

\author[Cao \& Pinsonneault]{
Lyra Cao\thanks{E-mail: \href{mailto:cao.lyra@gmail.com}{cao.lyra@gmail.com}} and Marc H. Pinsonneault
\\
Department of Astronomy, The Ohio State University, Columbus, OH 43210, USA\\
}

\date{Accepted 2022 September 16. Received 2022 September 11; in original form 2022 August 4}

\pubyear{2022}

\begin{document}
\label{firstpage}
\pagerange{\pageref{firstpage}--\pageref{lastpage}}
\maketitle

\begin{abstract}
We measure starspot filling fractions for 240 stars in the Pleiades and M67 open star clusters using APOGEE high-resolution H-band spectra. For this work we developed a modified spectroscopic pipeline which solves for starspot filling fraction and starspot temperature contrast. We exclude binary stars, finding that the large majority of binaries in these clusters (80\%) can be identified from Gaia DR3 and APOGEE criteria---important for field star applications. Our data agree well with independent activity proxies, indicating that this technique recovers real starspot signals. In the Pleiades, filling fractions saturate at a mean level of $0.248\pm0.005$ for active stars with a decline at slower rotation; we present fitting functions as a function of Rossby number. In M67, we recover low mean filling fractions of $0.030\pm0.008$ and $0.003\pm0.002$ for main sequence GK stars and evolved red giants respectively, confirming that the technique does not produce spurious spot signals in inactive stars. Starspots also modify the derived spectroscopic effective temperatures and convective overturn timescales. Effective temperatures for active stars are offset from inactive ones by $-109\pm11$ K, in agreement with the Pecaut \& Mamajek empirical scale. Starspot filling fractions at the level measured in active stars changes their inferred overturn timescale, which biases the derived threshold for saturation. Finally, we identify a population of stars statistically discrepant from mean activity-Rossby relations and present evidence that these are genuine departures from a Rossby scaling. Our technique is applicable to the full APOGEE catalog, with broad applications to stellar, galactic, and exoplanetary astrophysics.
\end{abstract}




\section{Introduction}\label{sec:introduction}
Magnetic fields are omnipresent in cool stars. The relationship between magnetism, rotation and convection is crucial for understanding stellar dynamos and the impact of magnetism on stellar structure and evolution. In bright and active stars, we can study the topology and strength of magnetic fields, or even image large starspot complexes. However, for the majority of targets, the direct measurement of surface field strengths is more challenging. Therefore studies of stellar activity have traditionally emphasized more indirect proxies, such as coronal or chromospheric activity, which can then be compared to stellar rotation as a function of stellar mass.

These chromospheric activity measurements were the first means by which astronomers have studied the stellar dynamo. Rich datasets of chromospheric activity has been measured in populations of stars with strong emission lines such as Ca II H\&K \citep[i.e. ][]{1966ApJ...144..695W, 1978ApJ...226..379W, 1980PASP...92..385V} and H$\alpha$. With a Ca II H\&K dataset \citet{1967ApJ...150..551K} discovered a strong correlation between chromospheric activity and convection. This was followed by the insight of \citet{1972ApJ...171..565S} that stellar magnetic activity declines with age according to an approximately inverse square root law.
Early on it was also discovered that coronal X-ray emission was correlated with rotation \citep{1981ApJ...248..279P, 1987ApJ...315..687M}.

Mass also matters; even at the same rotation rate, late type stars are more active than early type ones at the same rotation rate. A remarkable insight was the recognition that stars fall along a single sequence when activity is measured as a function of Rossby number---the ratio of the rotation period to the convective overturn timescale \citep{1978GApFD...9..241D, 1984ApJ...279..763N}. Activity proxies rise steeply at high Rossby number but tend to flatline at low Rossby number; this saturation phenomenon is observed for coronal emission \citep{2022ApJ...931...45N, 2011ApJ...743...48W, 2003A&A...397..147P} as well as chromospheric diagnostics \citep{2021A&A...656A.103F, 2018MNRAS.476..908F, 2010MNRAS.407..465J, 1993ApJS...85..315S}. Though it is a stellar evolutionary output, convective turnover timescale has traditionally been estimated empirically using these activity diagnostics \citep[e.g. ][]{1984ApJ...279..763N, 2003A&A...397..147P, 2011ApJ...743...48W}. Even as we work within the Rossby framework using activity proxies, there are a number of pressing questions.
Are stellar dynamos for fully convective stars and those for stars with radiative cores the same? Do other predicted events in stellar evolution, such as core-envelope decoupling, have a signature in starspots and activity? What causes the observed activity saturation: a maximum effective field strength at the surface, a complexification of the stellar magnetic field, or something else? And lastly, how do we map these activity proxies onto magnetic fields and modelling observables?

These and related questions about the nature of the stellar dynamo require a large and homogeneous magnetic dataset to answer. Direct magnetic mapping using Zeeman-Doppler Imaging is an approach which has contributed a valuable perspective to studies of the stellar dynamo; historically, however, this has been limited to very bright nearby samples where dedicated observation programs can be set up at high cadence and signal-to-noise \citep{2019ApJ...876..118S, 2014MNRAS.441.2361V}. Recent advances in applying magnetic effects to survey spectra with the Zeeman broadening technique have yielded larger datasets and exciting insights about the stellar dynamo in low mass \citep{2022A&A...662A..41R} and pre-main sequence stars \citep{2021ApJ...921...53L}. It appears to be possible to do magnetic measurements at low magnetic field strengths when carefully choosing Zeeman-sensitive lines \citep{2022A&A...662A..41R}, but doing whole-spectrum fitting may be limited to the most magnetic stars $\gtrsim$700 G, requiring that the Zeeman effect overwhelm rotational broadening \citep{2021ApJ...921...53L, 2020RNAAS...4..241H}.

There is another family of magnetic detections which involves fitting for the second temperature component in spectra; this method produces an observable ($f_{\mathrm{spot}}$) which can be fed directly into evolutionary models incorporating starspots \citep[e.g. ][]{2015ApJ...807..174S, 2022arXiv220803651J}. \citet{2016MNRAS.463.2494F} has applied an approach with spectral indices to obtain starspot measurements with the optical LAMOST survey for the Pleiades. However, these measurements are highly imprecise, likely because the contribution of spot flux is minimal in the optical; these measurements are complicated by the fact that starspot estimates from similar spectral indices such as TiO2 and TiO5 do not strongly correlate with each other \citep{2016MNRAS.463.2494F}.
Instead of optical wavelengths, \citet{2017ApJ...836..200G} demonstrated that precise starspot detections were possible with infrared spectra using the IGRINS instrument. The relative flux contribution by a cool component is higher in the infrared than in the optical, as the flux ratio approaches a constant on the Rayleigh-Jeans tail \citep{2017ApJ...836..200G, 1996AJ....111.2066W}. However, only a small number of spectroscopic starspots estimates are available with this method so far \citep{2022ApJ...925....5G, 2017ApJ...836..200G}.

In young stars, large starspots are required in order to reconcile theoretical models with observations; it may be crucial to understand the role of starspots and their impact on pre-main sequence evolution. Observations of these huge spot coverage fractions have been made \citep{2017ApJ...836..200G}, and they appear to satisfactorily explain age skew and radius inflation in isochrones \citep{2020ApJ...891...29S, 2017AJ....153..101S, 2016A&A...586A..52J, 2016A&A...593A..99F}. Models incorporating starspots and magnetic effects result in more consistent ages inferred with magnetic models compared to nonmagnetic ones \citep{2022ApJ...924...84C,2019ApJ...872..161D,2019ApJ...884...42S}; spotted models also allow lithium abundance variations to be explained as an ``activity spread'' in starspots rather than an unphysically large age spread \citep{2022MNRAS.513.5727B}. An observed spread in lithium can also be produced with a distribution of starspots, as was the insight of \citet{1993AJ....106.1059S}. The success of spotted and magnetic models in these and related problems motivates the development of a set of empirical detections of starspots for a large sample; this is a necessary step in calibrating models for precise stellar parameter recovery.

Another interesting avenue is in the recovery of starspot information from photometric variability data, which can potentially provide geometric and spot variability information when the mean starspot filling fraction is known; this avenue is very attractive due to the large number of precision light-curve data available in transit and transient surveys.
\citet{2021ApJ...916...66B} performed measurements of the mid-frequency continuum (MFC) by measuring the root-mean-square flicker of time-series photometry, finding a strong relationship between MFC flicker and rotation---observing that the activity saturation threshold was different from other activity measurements. \citet{2019ApJS..244...21S} also demonstrates a strong activity proxy using their {$S_{\mathrm{ph}}$} photometric flicker technique which is sensitive to starspot modulation. There has also been work to obtain starspot estimates by directly inferring starspot configurations from time-series photometry \citep{2020ApJ...893...67M}. However, the modulation of starspots across the stellar surface is strongly degenerate with inclination and surface distribution, and affected by spot lifetime effects---enough to concern interpretations of single-band photometry starspot inferences \citep[as pointed out by][]{2021AJ....162..123L, 2021AJ....162..124L}; published starspot estimates using optical photometry may also be insensitive to the small flux ratio of the cool spot component which would be visible in the infrared \citep{2017ApJ...836..200G, 1996AJ....111.2066W}. It appears that a potential solution to solving light curve degeneracies is multi-band time-series photometry \citep{2021AJ....162..124L}, or to affix the mean starspot filling fraction from a spectroscopic measurement and then fit for its variation through light curves \citep{2022ApJ...925....5G, 2017ApJ...836..200G}. It is also possible to look at the expected perturbation in the SED due to starspots \citep{1996AJ....111.2066W}, which will be possible to do precisely for large samples of stars with {\gaia} DR3 spectro-photometry. These rich time-series spectro-photometric datasets are valuable to the study of starspots and activity across large populations, and would benefit significantly from a large, homogeneous magnetic catalog.

The high-resolution H-band spectra of the Apache Point Observatory Galactic Evolution Experiment \citep[APOGEE, ][]{2017AJ....154...94M} is uniquely suited for providing starspot measurements---with its large sky footprint, high resolution infrared spectra, and detailed empirical linelist.
If starspot detections can be shown to be of sufficient precision with the APOGEE survey, the size of the accumulated infrared spectra would mean magnetic detections for hundreds of thousands of stars in DR17, and still more in Milky Way Mapper \citep{2017arXiv171103234K}; this would enable large populations of stars to be probed in activity, enabling for the first time a detailed and quantitative study of stellar dynamo physics, stellar evolution theory, and related fields in planet formation and habitability. In this paper we present the method and starspot detections on high-resolution near-infrared spectra from APOGEE Data Release 16, using the rich selection in open clusters to benchmark our technique.

We begin with two of the best-studied open clusters: the Pleiades and M67. Open clusters are ideal laboratories for testing stellar physics where a sequence of member stars at the same unique age can be identified. This yields stellar activity relations for coeval single stars with minimal contamination from field-age nonmembers. From the cluster locus on a color-magnitude diagram it is possible to remove binaries where the secondary contributes significant flux to the system. The rejection of binaries where the secondary contributes significant flux is important for a spectroscopic method which may see multiple temperature components; the most rapid rotators may also be interacting, and these populations of stars may require more careful analysis before their activity is interpreted on the same relation as single stars. With open clusters, it is possible to test stellar activity relations at distinct phases of stellar evolution. At young ages, stars are expected to be active and spotted; at old ages, this activity should decline to solar-like levels in solar-type stars.

In this work we present a method of inferring independent spectroscopic starspot measurements using our two-temperature method for open clusters, publish starspot constraint relations, and identify trends with activity.
In Section \ref{sec:methods} we describe the methodology for estimating starspot filling fractions and our experiment for estimating the fidelity of our technique. We describe the underlying structure of the pipeline (Section \ref{sec:apogeenir}), the construction of a two-temperature starspot grid (Section \ref{sec:gridconstruction}), the methods of binary rejection (Section \ref{sec:binaryrej}), the accuracy of our binary rejection routine (Section \ref{sec:binaryacc}), and the influence of starspots on the theoretical Rossby number (Section \ref{sec:rossbystarspots}). In Section \ref{sec:data} we indicate the data sources for the experiment. After computing starspot filling fractions we analyze these measurements and their relationships with activity in Section \ref{sec:analysis}. This includes a description of the data products and the relevance of binary rejection (Section \ref{sec:analysisfspot}). This is followed by an analysis of trends in starspots and magnetism with activity and rotation in the Pleiades (Section \ref{sec:spotcorrel}). We also test this technique on the old galactic open cluster M67 and discuss applications to solar-type and evolved stars (Section \ref{sec:discm67}). Synthesizing results from open clusters, we characterize the fidelity of this spectroscopic method in Section \ref{sec:accuracyprecision}. Finally, we discuss systematics between global stellar parameters recovered from our technique and from non-spotted catalogs in Section \ref{sec:systematics}. We discuss our results in Section \ref{sec:discussion} and summarize our findings in Section \ref{sec:conclusions}.

\section{Methods}\label{sec:methods}

\subsection{Fits to APOGEE High Resolution Near-IR spectra}\label{sec:apogeenir}

The ASPCAP \citep{2016AJ....151..144G} and IN-SYNC \citep{2014ApJ...794..125C} pipelines are powerful engines for inferring the properties of stars from spectra. However, ASPCAP does not include starspots in its fitting algorithm, so we have modified it to do so.

ASPCAP uses a least-squares minimization approach to spectral fitting. In addition to effective temperature and surface gravity, the pipeline also searches to ensure that the background atmosphere uses an internally consistent overall metal abundance [M/H], alpha-enhancement $[\alpha/Fe]$, carbon abundance [C/M], nitrogen abundance [N/M], and microturbulence. ASPCAP then solves for the best-fit parameters, including quality control flags that diagnose marginal or unacceptable fits. In a post-processing step, the key stellar parameters are placed on an absolute reference system, and individual element abundances are derived from subsets of lines.

The ASPCAP pipeline and linelists are the result of an enormous amount of coordinated work and have been verified on large populations of stars. We take advantage of this precise, existing machinery with the LEOPARD pipeline presented in this work, a starspot search grid for APOGEE high resolution ($R$$\sim$$22,500$) H-band (1.5–1.7 $\mu$m) spectra using empirical linelists capable of providing reliable detections of magnetism and activity on hundreds of thousands of stars. We use the {\sc{ferre}}\footnote{{\sc ferre} is publicly available from \url{http://hebe.as.utexas.edu/ferre}.} least-squares fitting spectra analysis code which has been applied with success to the APOGEE project. In the ASPCAP technique, {\sc ferre} spectral fits are done in an six-dimensional space for dwarfs: effective temperature ($T_{\mathrm{eff}}$), surface gravity (logg), $v \sin \, i$, microturbulence, [$\alpha$/Fe], and [M/H]; in the giants, [C/M] and [N/M] replace $v \sin \, i$ in the search grid for a seven-dimensional fit. Subsequent inferences in spectral windows and the use of the [$\alpha$/Fe] dimension allow the detailed estimation of abundances \citep{2016AJ....151..144G}.

The full eight-dimensional fitting procedure is expensive enough to make it computationally impractical to simply add additional degrees of freedom.  Instead, we take advantage of the fact that the C and N vectors are most important for highly chemically evolved giants, while our main targets of interest are main sequence stars, young stars, and less luminous giants. We can therefore replace these dimensions with ones that we use to characterize starspots. We approximate the impact of starspots by assuming that the spectra can be fit with a two-temperature solution. Although this is clearly a simplification of a more complex physical reality, it is motivated by clear evidence that cool spots cover a large fraction of the surface area of active stars. These spots induce large enough deviations in the spectral energy distribution to produce color anomalies; it is therefore a reasonable hypothesis that a similar signature might be detected via spectroscopy. We adopt a two-temperature model where the observed spectral flux is composed of hot and cool components with a unique temperature contrast and a spot filling fraction, which decide the temperature spacing between the cool and the hot component and the amount of cool component to add to the flux, respectively. The hot and cool component fluxes are summed together along one dimension of a spectroscopic search grid. The signal of two-temperature models is in the continuum-normalized lines which are perturbed by the excess flux from a cool component. For local open clusters in particular, variations in [$\alpha$/Fe], [C/M], and [N/M] are not expected to be as important in the spectra, so we replace these three dimensions with two starspot dimensions: $f_{\mathrm{spot}}$ and $x_{\mathrm{spot}}$. This results in a seven-dimensional fit over $T_{\mathrm{eff}}$, logg, $v \sin \, i$, microturbulence, [M/H],  $f_{\mathrm{spot}}$, and $x_{\mathrm{spot}}$.

The use of near-infrared spectra for detecting two temperature components has been previously explored by \citet{2017ApJ...836..200G} on the echelle spectra of the IGRINS spectrograph; to detect the impact of binary companions in APOGEE on the composite stellar spectrum, \citet{2018MNRAS.473.5043E} applied a two-temperature binary solution using a similar search grid. In the latter case, the additional degree of freedom was the light from a companion star on the same isochrone as the primary.

In this work we process one visit spectrum per star, as the spot filling fraction can change, or be modulated by rotation, on timescales that can be shorter than the mean elapsed time between visits. The typical S/N of a visit spectrum in our Pleiades single-star sample is high with a mean value of $158$/pixel. APOGEE visit spectra are available for many stars and may sample variations in activity due to intrinsic processes on the star over long timescales. A full analysis of available APOGEE visit spectra is planned for future work.

\subsection{Grid Construction \label{sec:gridconstruction}}
To construct the starspot grid, we populate a 7-D grid with synthetic spectra by using model atmospheres from {\sc{phoenix}} \citep{2013A&A...553A...6H} with APOGEE linelists \citep{2021AJ....161..254S} in the publicly available spectral synthesis program SYNSPEC \citep{2021arXiv210402829H}. The choice of {\sc{phoenix}} rather than the {\sc{MARCS}} model atmospheres was made in this work because of the finer resolution (in steps of $100$ K, instead of $250$ K) of the atmosphere grid in $T_{\mathrm{eff}}$ between 4000–7000 K; in our exploration of field stars, we will explore using {\sc{MARCS}}, which has a significantly larger metallicity range in their models. We convolve these with the APOGEE combined line spread function and a set of microturbulence parameters identical to the treatment in ASPCAP for dwarfs.

To generate synthetic two-temperature spectra we relate the components $T_{\mathrm{spot}}$ and $T_{\mathrm{amb}}$ with the temperature contrast $T_{\mathrm{spot}} = x_{\mathrm{spot}} T_{\mathrm{amb}}$ following the approach of \citet{2015ApJ...807..174S}:
\begin{equation}\label{eqn:twoteff}
T_{\mathrm{eff}} = T_{\mathrm{amb}} \left( 1 - f_{\mathrm{spot}} + f_{\mathrm{spot}} x_{\mathrm{spot}}^4 \right)^{\frac{1}{4}} .
\end{equation}
With a choice of $T_{\mathrm{eff}}$, $f_{\mathrm{spot}}$, and $x_{\mathrm{spot}}$ we can uniquely identify a pair of temperature components $T_{\mathrm{amb}}$ and $T_{\mathrm{spot}}$. The temperature components are then summed in an area-weighted fashion over wavelength which is flux-conserving and consistent by construction:
\begin{equation}\label{eqn:fluxtwoteff}
B \left( T_{\mathrm{eff}} , \lambda \right) = f_{\mathrm{spot}} B \left( T_{\mathrm{spot}} , \lambda \right) + \left( 1 - f_{\mathrm{spot}}\right) B \left( T_{\mathrm{amb}} , \lambda \right)
\end{equation}
We use the same values of logg, $v \sin \, i$, [M/H] for both the spotted and ambient components. After identifying the two unique temperatures for a triple in $T_{\mathrm{eff}}$, $f_{\mathrm{spot}}$, and $x_{\mathrm{spot}}$ we interpolate on the underlying synthetic spectra fluxes across $T_{\mathrm{eff}}$ with a cubic spline to obtain the cool and hot flux components to be used in the area-weighted flux sum. For spectroscopic fitting we perform quadratic spline interpolations across the grid as part of the {\sc{ferre}} procedure. These higher-order interpolations on the fluxes appear to be more accurate and physically appropriate than linearly interpolating on model atmosphere structures \citep{2013MNRAS.430.3285M}.

The current set of models are defined on PHOENIX models for effective temperature between $3300$–$12000$ K, logg between $2.50$–$5.50$, [Fe/H] between $-0.5$–$+0.5$, and vsini's at logarithmic steps between $1.5$–$96$ km/s. A spot filling fraction is defined between $0$–$1$ in increments of $0.05$ and a spot temperature contrast between $0.8$–$1.0$ in steps of $0.05$. Since a spot temperature contrast is defined for $x_{\mathrm{spot}}=$ $0.8$–$1$, the effective temperature is bounded at both low and high temperature. At $3300$ K a model of $2640$ K must be defined for a spot contrast of $0.8$; at the other extreme, $12000$ K, a model of $15000$ K is required for $x_{\mathrm{spot}} = 0.8$. To avoid grid edge issues at the low temperature end we set a {\gaia} $G_{\mathrm{BP}}-G_{\mathrm{RP}}$ restriction of 0.5–2.5 in this work; this cut, spanning roughly the spectral types F2–M3 and the effective temperatures $3400$–$6800$ K, is imposed to limit the complexity of accounting for mid-to-late M dwarf linelists and the interpretation of our solutions on hot stars at present.

We find that our starspot technique can use a single grid for all stars by using the unconstrained optimization by quadratic approximation (UOBYQA) method as implemented in {\sc{ferre}} \citep{powell2002uobyqa}. Other numerical techniques often find difficulty converging with such a large $T_{\mathrm{eff}}$ range because of physical non-uniqueness of stellar solutions in spectra, a problem which has led to the development of multiple subgrids in projects like ASPCAP. However, we find robust agreement with non-spotted stellar parameter solutions and a strong improvement in reduced $\chi^2$ for active, young stars in our two-temperature solutions in this paper. Any convergence failures which may result from this nonlinear search are rejected.

The reported errorbars from {\sc{ferre}} are obtained by inverting the curvature matrix assuming all dimensions are independent. Since there may be covariances between many dimensions, the errorbars may in theory be overestimated relative to a formalism considering these covariances. However, systematics dominate over the reported random errors from these high resolution spectra. Possible systematics include mismatches to linelists, unresolved binaries, and additional complexities in starspot structure. Reported errorbars significantly smaller than the size of a single grid spacing may be underestimated for individual stars because of systematics associated with gridding \citep{2015ApJ...812..128C}. Since the two-temperature method involves using linear weighted sums of spectra, this concern is not likely to be significant for the interpretation of starspot measurements.

In addition to starspots, the 7-D fit in this work provides the other jointly-fit stellar parameters such as $T_{\mathrm{eff}}$, logg, [M/H], and $v \sin \, i$. As these parameters can be compared with literature values, we explore the systematics in detail in Section \ref{sec:systematics}. It's important to note that even though we have removed dimensions such as [$\alpha$/Fe], [C/M], and [N/M], we still obtain elemental abundances by doing fits on spectra in spectral windows.

\subsection{Binary Rejection}\label{sec:binaryrej}
The detection of a starspot signature with spectra is confounded by the presence of a companion---as binaries can be physical two-temperature spectroscopic solutions. Chance background blends are also a potential source of contamination on the same APOGEE fiber as the flux contribution from another star on the fiber may erroneously appear as a high spot filling fraction. Though it may be useful to investigate binary populations with this technique, which will need a careful accounting for the additive effects of binarity and activity on spectra, we discard binaries here to test activity relations in the single star population.

Open clusters are particularly useful for efficiently rejecting photometric binaries after identifying the cluster sequence; this removes sources where the secondary contributes a significant fraction of the observed flux, which can interfere with our two-temperature spectroscopic solution. For field or young populations where identifying photometric binaries is impossible, we propose a set of additional kinematic binary discrimination diagnostics to reduce the amount of binary contamination. The purity of this binary rejection procedure can then be tested on the Pleiades and M67, which gives a first estimate for the binary contamination of a large field starspot catalog.

\subsubsection{Photometric Binary Cut}\label{sec:photbinary}

To perform a photometric binary cut, we first identify a single star sequence defined as the 75th percentile running mean in 2MASS K-band magnitude. We then find an empirical fourth-order polynomial fit as an empirical single-star isochrone, and use it to identify stars brighter than this fiducial by $0.25$ mag; this procedure is slightly more conservative than the field star binary exclusion cut proposed for the \textit{Kepler} fields by \citet{2019ApJ...871..174S}. This empirical isochrone avoids known isochrone skews and trends for the active and spotted low-mass stars \citep{2017AJ....153..101S}. We now turn to other independent diagnostics for the presence of a companion.

\subsubsection{APOGEE \& {\gaia} DR3 Radial Velocity Cut}\label{sec:rvvar}

The radial velocity fit information from APOGEE and {\gaia} {\it Radial Velocity Spectrometer} (RVS) spectra provides independent information which is sensitive to the astrometric motions of close binaries and stars with massive companions. These scatter in these RV measurements provides indications of spectroscopic binarity \citep{2022arXiv220605902K}. Due to the different systems on which these RV measurements are made and reported, we analyze these criteria separately.

The APOGEE \verb|VSCATTER| parameter represents the variation in the radial velocity measurements of successive visits; this value is defined for a large number of Pleiads since it is a well-sampled nearby open cluster in APOGEE. To select a sample of non-RV variable sources, we choose a cut of 1 km/s as our threshold for binarity; this is the same value used in \citet{2020AJ....160..120J}.

The RV variability of {\gaia} RVS spectra is well-documented in the RVS validation paper \citep{2022arXiv220605902K}. They advocate a workflow for identifying RV variable stars by screening in stars for which \verb|rv_chisq_pvalue|, the likelihood of a constant radial velocity measurement over time, is low. Additionally, they combine this with \verb|rv_renormalized_gof|, a renormalized unit-weight error parameter from the RV time series which is much larger than unity for variable sources. We adopt this criteria for identifying binarity as written: \verb|rv_chisq_pvalue|$\leq 0.01$ \& \verb|rv_renormalized_gof|$>4$ \& \verb|rv_nb_transits|$\geq 10$ \& $3900 \leq$\verb|rv_template_teff|$\leq 8000$.

While these kinematic cuts may in some cases involve companions without a significant flux contribution, for this work we aim for a sample with a high purity of single stars to characterize the behavior of our starspot solution.

\subsubsection{{\gaia} RUWE Cut}\label{sec:ruwecut}
The Re-normalized Unit Weight Error (\verb|RUWE|) parameter in {\gaia} represents the square root of the reduced $\chi ^2$ statistic for an astrometric solution, corrected for photometric correlations \citep{2020ApJ...894..115P, 2018A&A...616A...2L}. Small departures above a value of unity are strongly correlated with the presence of a binary or tertiary companion \citep{2021ApJ...907L..33S}, and appear to be useful in determining systems with semimajor axes of $0.1$–$10$ AU for systems within $2$ kpc \citep{2020MNRAS.496.1922B}. Factors affecting the point spread function of the star have also been speculated to increase the \verb|RUWE| parameter, which may due to a chance blend or a semi-resolved binary \citep{2020MNRAS.496.1922B}. As pointed out by \citet{2022A&A...657A...7K}, {\gaia} \verb|RUWE| is a statistical measurement and appears also to be correlated to proper motion anomaly from independent binarity analyses from Hipparcos.

One convention in the literature is to use a criteria of \verb|RUWE|$>1.4$ to indicate binarity \citep[e.g. ][]{2022arXiv220605989B, 2022A&A...657A...7K}. However, as \citet{2022ApJ...931...45N} indicates, the choice of \verb|RUWE|$>1.2$ is also used as a more conservative binary criterion \citep{2020ApJ...894..115P}.
Since the purity of our single-star sequence is essential to the interpretation of our starspot parameter, we adopt the {{\gaia} \verb|RUWE|} cut of 1.2.

\subsubsection{Gaia Multiple Source Cut}\label{sec:gaiamultiple}
There is a possibility that multiple sources can be identified in {\gaia} overlapping on the same 2 arcsecond diameter APOGEE fiber, including wide binaries and blended sources. To mitigate this possibility, we also use {\gaia} DR3 to detect multiples of named sources within a conservative distance of 3 arcsec. Any source detected will be flagged as a multiple source in a field like the Pleiades; the choice of 3 arcseconds is larger than the diameter of APOGEE to remove an excess of possible blended sources. Even if the sources are on the same fiber, there is also a possibility that the faint source does not contribute enough source for the solution to be significantly affected in APOGEE. For instance, a neighboring source that is fainter by $\geq 5$ magnitudes in H-band contributes less than $1\%$ of flux to the stellar spectra. We remove these stars in this paper, but it is prudent to explore these limits and their impact on the starspot solution in future work.

\subsubsection{M67 External Radial Velocity Cut}\label{sec:m67extcut}
M67 is significantly more crowded than the Pleiades, so we do not use multiple source detections in {\gaia} as a binary cut; instead, we use a comprehensive external binarity catalog to remove stars. \citet{2021AJ....161..190G} publishes the results of a careful RV variability analysis with a 45 year measurement baseline, which they estimate is complete to $\sim$91\% down to $V=16.5$ and $>$99.5\% to $V<15.5$; we expect a high completeness of our sample, since all our stars are brighter than $V=16.5$ with $\sim$91\% brighter than $V=15.5$. However, there is likely an underlying binary contamination level, as long period ($P > 10^4$ d), high eccentricity, or low mass ratio systems can be beyond their detection limits \citep{2021AJ....161..190G}.

\begin{figure*}
\includegraphics[trim={0cm 0cm 0cm 0cm},width=\textwidth]{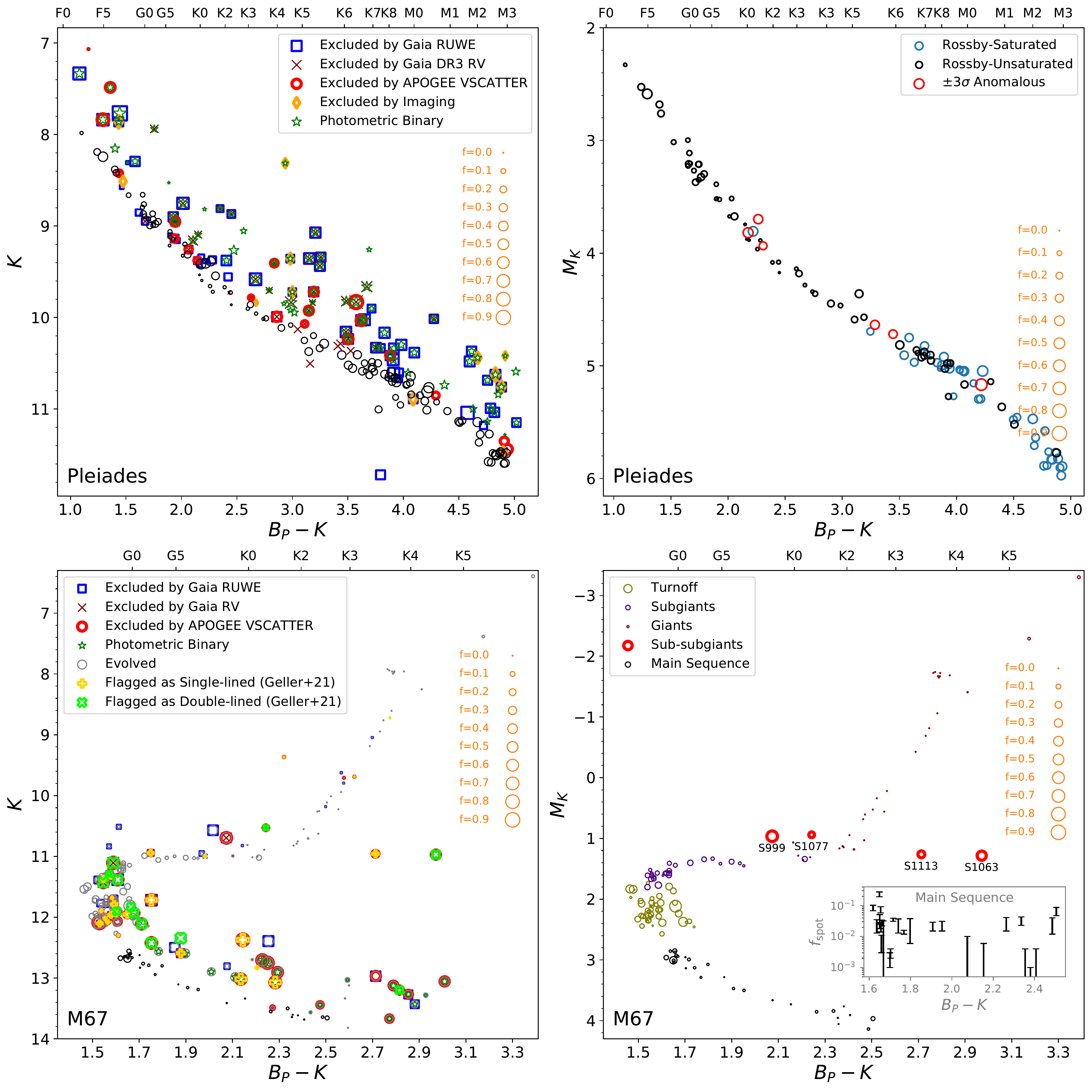}
\vspace{-1.5em}
\caption{\label{fig:SP2aI_CMDs_big} Color magnitude diagrams for the Pleiades (top) and M67 (bottom). Left: Binary rejection techniques are shown as different symbols scaled by the size of the starspot filling fraction inferred in Section \ref{sec:analysisfspot}. Stars which have multiple classifications show up as different symbols of roughly the same size on top of each other. Stars detected by any binary flag are rejected from the subsequent analysis. Right: the single-star diagrams are shown for each cluster. On the top is the CMD for the rotationally saturated, unsaturated, and flagged anomalous stars in the Pleiades, these designations explained further in Section \ref{sec:spotrossby}. On the bottom is M67 with each evolutionary state identified for a sequence which passes each of our binarity cuts, as well as the interacting sub-subgiants added back in for effect. The inset plot on the bottom right panel shows the spot distribution for the M67 main sequence GK stars.}
\end{figure*}

\subsection{Accuracy of Binary Removal}\label{sec:binaryacc}

In Figure \ref{fig:SP2aI_CMDs_big} each of the binary rejection techniques is shown working in concert for the same stars; if multiple approaches flag binarity, the same approximate size symbols are shown overlapping on top of each other. The size of the points is proportional to the starspot filling fraction which is discussed in Section \ref{sec:analysisfspot}.

Out of 214 stars in the Pleiades with detected rotation periods in APOGEE DR16 with $0.5 < G_{\mathrm{BP}}-G_{\mathrm{RP}} < 2.5$, 109 were flagged as single stars and 105 were flagged as binaries. This leads to a binary rejection fraction of $49\%$. This is consistent with the low end of the binary ratio found in the recent work by \citet{2022AJ....163..113M}, which found a binarity fraction between $0.54\pm0.11$ and $0.70\pm0.14$ in the Pleiades. The photometric rejection criterion is effective at flagging binaries for 72 of the 105 binaries, which is itself complete to $69\%$ of our final binarity criteria. Furthermore, 89 of the 105 binaries were flagged with criteria from {\gaia} RV variation, APOGEE RV scatter, {\gaia} \verb|RUWE|, or {\gaia} multiple source detections, meaning that $\sim$85\% of binaries are rejected without the photometric rejection constraint; this is an estimate for the completeness of these criteria across the sky.

The performance of each non-photometric rejection criteria can also be analyzed. The APOGEE \verb|VSCATTER| parameter is a suitable cut for 98 of the 214 stars, and out of these 98, just 19 ($19\%$ of stars for which \verb|VSCATTER| is defined, $9\%$ of sample) were excluded. The {\gaia} RV variation criteria is defined for 119 out of 214 stars, and excludes 40 of them ($34\%$ of stars for which the {\gaia} RVS parameters are defined, $19\%$ of the sample). {\gaia} \verb|RUWE| was defined for 209 of the stars, excluding 60 of them ($28\%$ of stars in our sample). Finally, the multiple source criteria was effective at rejecting 12 out of 214 sources (5\%). In field or pre-main sequence regions where photometric rejection is not possible, the relative impact of each of these binary rejection filters may change. Any remaining binaries can be rejected with a more complete astrometric solution or a crossmatch to a dedicated binary catalog. Additionally, it is possible that a two-temperature solution to binaries themselves can be used as a filter; rather than using the temperature contrast parameter for starspots, this would involve building a binary grid similar to \citet{2018MNRAS.473.5043E} for flagging binaries.

We perform the \verb|RUWE|, {\gaia} RV variation, APOGEE \verb|VSCATTER|, and photometric cuts in M67. This yields a clean single cluster sequence for the solar-like main sequence stars. In M67 the main difficulty is that photometric binary exclusion is not trivial to do on the main sequence turnoff. Fortunately, there are long-term dedicated binary studies in M67. We take advantage of this information by including the flagged single-lined and double-lined spectroscopic binaries from \citet{2021AJ....161..190G} as an additional binarity flag.

It is important to note that the two-temperature starspot technique should be confounded only in binaries where the secondary provides a significant flux contribution in the infrared. On the single-star sequence these starspot detections should not be suspect even if stars have kinematic motions which correspond to a companion, as they contribute minimal flux.

\subsection{Rossby Numbers with Starspots}\label{sec:rossbystarspots}
Rossby number, the ratio of the rotation period and the convective turnover timescale $\mathrm{Ro} = P_{\mathrm{rot}}/\tau_{\mathrm{CZ}}$, is often used as a diagnostic for the efficiency of the dynamo mechanism. \citet{1978GApFD...9..241D} first reasoned that the strength of the dynamo should be related to the Rossby number from a theoretical perspective. \citet{1984ApJ...279..763N} then observed a strong correlation between Rossby number and chromospheric activity, reducing scatter significantly in activity. Subsequent studies in chromospheric, coronal, and magnetic proxies have validated the practice of using Rossby numbers to parameterize the stellar dynamo. Convective turnover timescales are usually used as either an empirical fit parameter to reduce scatter in the activity---rotation relation \citep{2018MNRAS.479.2351W, 2011ApJ...743...48W, 2003A&A...397..147P, 1994A&A...292..191S, 1984ApJ...279..763N} or as an theoretical output of a stellar evolution code \citep{2017A&A...605A.102C, 2010A&A...510A..46L, 1996ApJ...457..340K}.
In this paper, we argue for $\tau_{\mathrm{CZ}}$ as a theoretical output rather than an empirical constraint; this is because even clusters and field populations may have a spread in activity and rotation at a given age, and some of these stars may be going through changes in their convective turnover timescales as a result of stellar evolution.

Active stars are more likely to support strong magnetic fields on their surfaces, causing structural changes in stellar parameters, including the convective turnover timescale. The empirical approach to estimating $\tau_{\mathrm{CZ}}$  can result in increased scatter in the Rossby number due to an inappropriate assignment of $\tau_{\mathrm{CZ}}$, and a systematic bias in Rossby number due to the non-magnetic treatment (a differential bias in the active stars, for which the structural effect is most important). With our measurements of spectroscopic starspot filling fractions it is now possible to self-consistently infer theoretical convective turnover timescales with a SPOTS \citep{2020ApJ...891...29S} isochrone, matching the starspot parameter between our observations and appropriate theoretical models. There are two differences between these convective turnover timescales and other commonly used prescriptions in the literature.

\citet{2020ApJ...891...29S} define the convective overturn timescale as the velocity divided by the pressure scale height at the point where the distance to the base equals the local pressure scale height. This differs from the traditional local approximation, where the pressure scale height is evaluated at the base of the convection zone, and the velocity is measured at that distance above the base. Because the pressure scale height diverges for fully convective stars, the traditional measure is ill-posed, while the \citet{2020ApJ...891...29S} definition is continuous. This is particularly advantageous for studying the stellar dynamo in the regime close to the fully convective boundary.

\begin{figure}
\includegraphics[trim={0cm 0cm 0cm 0cm},width=\columnwidth]{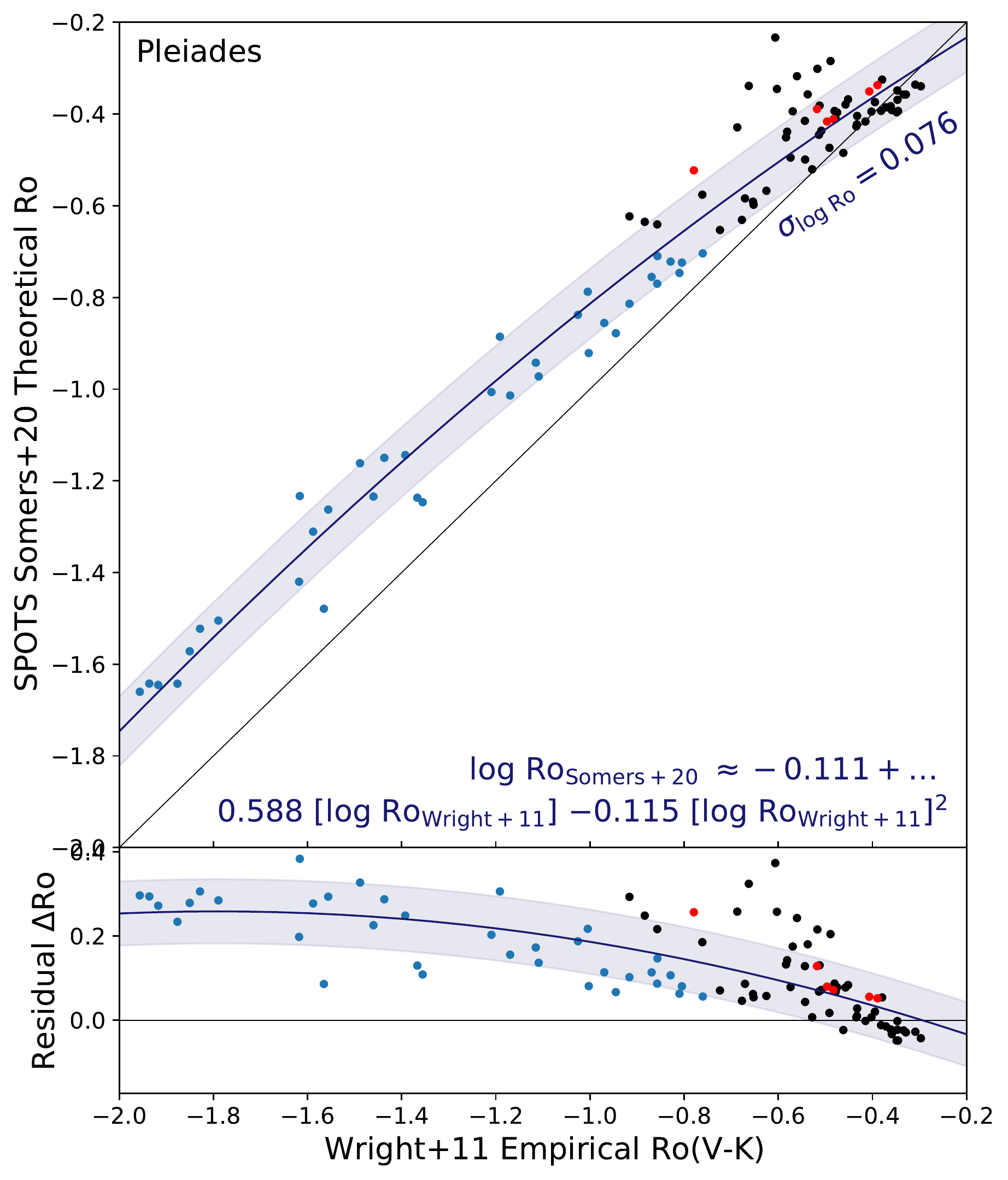}
\vspace{-1.5em}
\caption{\label{fig:SP2aI_Ro_Ro_big} Fit between Rossby numbers computed using $\tau_{\mathrm{CZ}}$ from the SPOTS grids \citep{2020ApJ...891...29S} and the empirical $\tau_{\mathrm{CZ}}$ of \citet{2011ApJ...743...48W} in the Pleiades single star sample. Black and blue symbols correspond to Rossby-unsaturated and saturated stars respectively, while red symbols indicate $3 \sigma$ anomalously active stars from Section \ref{sec:spotrossby}. A second order polynomial fit to the offset is shown between the theoretical and empirical Rossby numbers.}
\end{figure}

The SPOTS models also include the structural effect of starspots on the global stellar properties, including $T_{\mathrm{eff}}$ and $\tau_{\mathrm{CZ}}$. We show the differences between an empirical Rossby number from \citet{2011ApJ...743...48W} and Rossby numbers derived from the SPOTS models in Figure \ref{fig:SP2aI_Ro_Ro_big}. For most of the Rossby-unsaturated stars in black, there is no appreciable systematic; however, the Rossby-saturated stars in blue appear to systematically biased at the level of $- 0.2$ dex in models not incorporating the structural effects of starspots.
This is a natural consequence of the feedback between starspots and stellar structure; inactive stars have thermal structures and overturn timescales similar to unspotted models, while active stars are inflated in radius and different.
We suggest that the growing population of anomalously inactive M-dwarf rapid rotators identified from studies such as \citet{2022AJ....163..257A, 2017ApJ...834...85N, 2009ApJ...693.1283W} may be partially explained by the structural effect of starspots on their convective overturn timescales; if these stars show evidence of spots on their surface at the level expected from our analysis, the non-spotted Rossby numbers may be underestimated by up to $\sim$$0.2$–$0.3$ dex.

Finally, the global properties of the fit, including the location of the break at saturation, are shifted as a result of the starspot analysis by $0.05$–$0.2$ dex. We address this in Appendix \ref{sec:appendixrossby}.

\section{Data}\label{sec:data}

We propose an experiment which will test the accuracy and precision of our technique. Young stars are known to be active and rapidly rotating, with evidence of X-ray and UV excesses and chromospheric activity. On the other hand, old solar-type stars are known to be relatively slow-rotating and inactive. Identifying a well-studied young and old cluster will quantify the accuracy of starspot extraction; the precision will be estimated by looking at the variation in starspot filling fraction in domains where trends in starspot variation are minimized. Once the promise of the spectroscopic starspot method is established, we can explore its unique properties and its correlation to other activity diagnostics.

The 16th Data Release \citep{2020ApJS..249....3A,2020AJ....160..120J} of the Apache Point Observatory Galactic Evolution Experiment \citep[APOGEE, ][]{2017AJ....154...94M} reports a cumulative set of high resolution ($R\sim22,500$) reduced H-band (1.5–1.7 $\mu$m) spectra for 437,445 stars with spatial coverage in both the northern and southern hemispheres. In APOGEE DR16 there were also programs to observe representative open clusters, with M67 serving as a calibration cluster \citep{2021AJ....162..302B}. This targeting program provided near-infrared spectra of hundreds of stars per cluster, sampling clusters orders of magnitude better than what would have serendipitously be observed across the sky. These richly-sampled clusters are a key to understanding magnetism, activity, and the stellar dynamo.

Starting from published membership lists we obtain members from well-studied open clusters in the public DR16 dataset. These include the Pleiades \citep[$\sim$125 Myr;][]{1998ApJ...499L.199S} and M67 \citep[$\sim$4.3 Gyr;][]{1998ApJ...504L..91R}. These clusters are well-sampled by APOGEE DR16 and represent template active and old systems, respectively. We obtain kinematics and uniform photometry by crossmatching these lists to {\gaia} and 2MASS.

\subsection{Case Studies: Comparisons of $f_{\mathrm{spot}}$ with the literature}\label{sec:casestudies}

There are astrophysical case studies in the literature where Zeeman-Doppler imaging or spectral fitting has been published, and we compare our results with some example cases. These provide valuable tests for this machinery as these are independent measurements using external spectra. For this purpose we include well-studied stars from the literature to demonstrate two-temperature fits in active stars. The weak-lined T Tauri star LkCa 4 is a member of the Taurus star forming region which has been characterized previously in both accretion \citep{2001ApJ...556..265W} and activity \citep{2014MNRAS.444.3220D, 2017ApJ...836..200G}, and its literature starspot measurement is a test of the concordance of our technique on a heavily spotted young star in Section \ref{sec:analysislkca4}. An exemplar of the Rossby-saturated low mass regime, the Pleiad HII 296 has a published magnetic field measurement made with ZDI \citep{2016MNRAS.457..580F}, which we compare to our starspot measurement in Section \ref{sec:analysishii296}.

\subsection{Pleiades}\label{sec:datapleiades}
The Pleiades is one of the best-studied nearby open clusters, with a near-solar metallicity \citep[$\sim$0.03$\pm$0.02$\pm$0.05 dex;][]{2009AJ....138.1292S}. Lithium age dating is the most accurate measure for young open clusters. The initial estimate of 125$\pm8$ Myr \citep{1998ApJ...499L.199S} has been updated with new models to 112$\pm5$ Myr \citep{2015ApJ...813..108D}. There is a spread of as large as a factor of about 50 in the lithium abundance at fixed effective temperature in cool stars, implying a real spread in the destruction of lithium at the same mass, composition, and age \citep{2018A&A...613A..63B, 1993AJ....106.1059S, 1993ApJS...85..315S}. It has been suggested that this lithium spread may be the result of starspots and inflated stellar structure \citep{2014ApJ...790...72S, 1993AJ....106.1059S}; there has also been a recent suggestion that this spread could be caused by variations in convective efficiency as a result of rotation \citep{2021A&A...654A.146C}. For both mechanisms, the physical process is a supression of the lithium depletion during the pre-MS by an amount that varies from star to star. \citet{2003AJ....126..833S} found that the Pleiades K dwarfs were anomalously blue in their spectral energy distributions, and that this approximately correlated with $v \sin \, i$. All of these point towards a need for a consistent starspot interpretation of the Pleiades---\citet{2016MNRAS.463.2494F} found with a model of spectral indices in molecular TiO that low mass stars were spotted at the level of $\sim$$30\%$; unfortunately, these measurements had a characteristically large scatter.

Stellar rotation in the Pleiades has been extensively studied in the literature. The pioneering study of \citet{1987ApJ...318..337S} measured rotation velocities in a sample of unprecedented size and quality, finding that there were both slow and rapid rotators in the Pleiades with a strongly mass dependent pattern. With the advent of time domain surveys, the emphasis shifted to rotation period samples, which can obtain more precise measurements for slow rotators. Our source for rotation periods for the Pleiades are from targeted observations in K2 \citep{2016AJ....152..113R}. These observations reveal a distribution of multi-periodic light curves with different morphologies, which appear to be the result of rotational modulation by starspots or latitudinal differential rotation, with a distinct slow and rapid sequence in the higher mass stars \citep{2016AJ....152..114R, 2016AJ....152..115S}. To construct our sample, we require that each star have a rotation period; we begin by using the member list from \citet{2016AJ....152..113R}.

We collect X-ray luminosities using the published Pleiades {\it ROSAT} catalogs of \citet{1999A&A...341..751M} and \citet{1994ApJS...91..625S}, and the {\it Einstein} data from \citet{1990ApJ...348..557M}. There were also some serendipitous X-ray measurements from the {\it XMM–Newton} observatory, but owing to the large fraction of binaries and low number of single stars in the detections, we do not include these sources in this work \citep{2003MNRAS.345..714B}.

In the ultraviolet, {\it GALEX} data exist for the Pleiades \citep{2009PASP..121..450B, 2017ApJS..230...24B}. However these published catalogs use fields which largely avoid the center of the cluster---crowding and source contamination are concerns in the UV data, contributing to a very small number of identified members with UV detections. There has also been a number of studies to systematically study the Pleiades in the UV \citep[e.g. ][]{2003ApJ...589..347G} but with pixel scales ($\sim$15 arcsec) much larger than the APOGEE 2 arcsec fibers.

We collect chromospheric activity using published $R^{\prime}$ values, which indicate the fraction of excess flux emitted in lines such as Ca II IRT or H$\alpha$ to the total flux of the star; more recently, a mapping between emission line equivalent width and emission flux has made it possible to obtain these activity indicators from optical spectra. In the Pleiades, we include $R^{\prime}_{\mathrm{H}\alpha}$ and $R^{\prime}_{\mathrm{IRT}}$ from \citet{1993ApJS...85..315S}. To expand this sample to lower masses we also include $R^{\prime}_{\mathrm{H}\alpha}$ and $R^{\prime}_{\mathrm{Ca \; K}}$ inferred from equivalent widths in LAMOST spectra \citep{2018MNRAS.476..908F}. We make two assumptions to calibrate the spectroscopic activity measurements onto the photometric ones: we assume that the locus of activity variation is the same for stars in common and that the Ca emission line measurements are well-correlated (Appendix \ref{sec:appendixchromospheric}). We also obtained Ca IRT measurements inferred from {\gaia} DR3 \citep{2022arXiv220605766L}; however the measurements largely overlap with the \citet{1993ApJS...85..315S} sample, with just one reported $R^{\prime}_{\mathrm{IRT}}$ measurement of a Rossby-saturated star; thus the {\gaia} Ca IRT estimates do not replace the LAMOST Ca analysis in the Pleiades, and we have omitted it.


\subsection{M67}\label{sec:datam67}
The open cluster M67 has played a major role in testing the theory of stellar structure and evolution. It is close to the Sun in composition and age \citep[$4$ Gyr and {[Fe/H]} = $0$ respectively;][]{2008A&A...484..609Y, 2018ApJ...857...14S}, so M67 has been crucial for the solar-stellar connection in astrophysics. This has included studies of Li depletion \citep{1999AJ....117..330J, 2012A&A...541A.150P}; activity in solar analogs \citep{2006ApJ...651..444G}; and rotation periods at solar age \citep{2016ApJ...823...16B}. The APOGEE survey obtained a deep sample of spectra for M67 members \citep[see][]{2019ApJ...874...97S, 2021AJ....162..302B}, which mapped out surface abundance changes from gravitational settling from the lower main sequence to the turnoff and the giant branch. The cluster is also well-studied for binary populations \citep{2021AJ....161..190G}. For our purposes, M67 is a key calibration system. We expect sun-like activity levels in solar analogs in the cluster, so we can use M67 to ensure that our method does not predict spurious spot signals in inactive stars. The exceptions to the general pattern of inactivity on the main sequence are binary stars impacted by current or past tidal interactions. These include the two well studied interacting sub-subgiants, S1063 and 1113. These active stars are heavily spotted \citep{2022ApJ...925....5G}, and we discuss our derived starspot properties for them in Section \ref{sec:analysiss1063}. Because the cluster has already been analyzed for binarity, we can also test our ability to separate starspot signals from binary ones; this is essential for investigating field populations in the upcoming catalog.

We collect published M67 members by crossmatching the union of {\gaia} kinematic membership lists \citep{2018ApJ...869....9G, 2019A&A...627A.119C} with {\gaia} DR3, APOGEE DR16, and the comprehensive radial velocity study of \citet{2021AJ....161..190G}.

\section{Analysis}\label{sec:analysis}

\subsection{Outcome of $f_{\mathrm{spot}}$ fits}\label{sec:analysisfspot}
We ran our pipeline over the literature assigned members with an APOGEE spectrum, including 214 stars in the Pleiades (Section \ref{sec:datapleiades}) and 227 stars in M67 (Section \ref{sec:datam67}). This yielded spectroscopic inferences of $T_{\mathrm{eff}}$, logg, $v \sin \, i$, microturbulence, [M/H], $f_{\mathrm{spot}}$, and $x_{\mathrm{spot}}$.

\begin{figure}
\includegraphics[trim={0cm 0cm 0cm 0cm},width=\columnwidth]{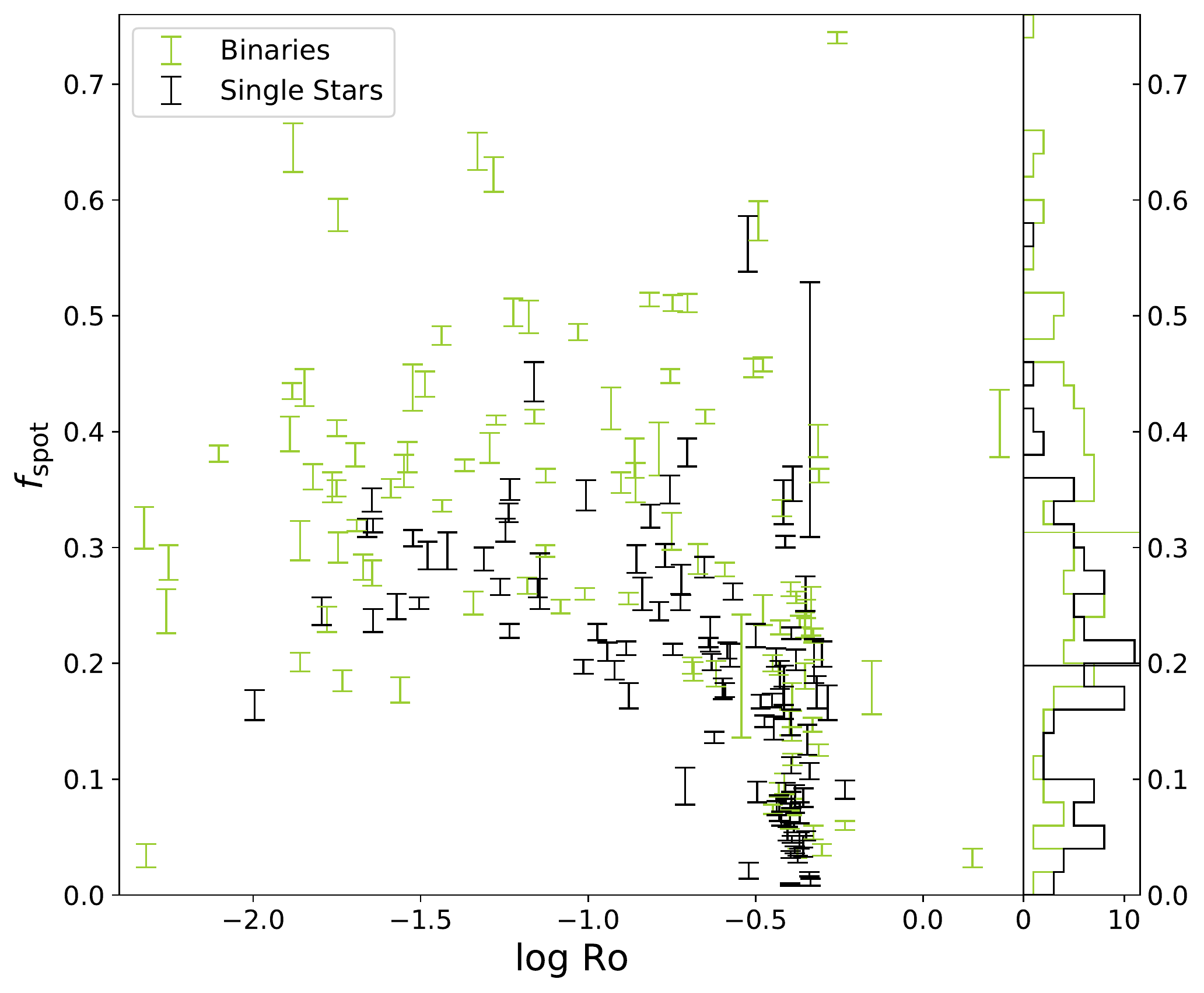}
\vspace{-1.5em}
\caption{\label{fig:SP2aI_fspot_Ro_binaries_big} Rossby Diagram with binaries labelled, showing the need for accurate binary rejection as the binary line lies above the single star line by about 0.1 in $f_{\mathrm{spot}}$ for the rapid rotators. Black symbols indicate the single stars while light green symbols indicate the binaries. Side: a histogram of the binaries and the single stars along with an vertical line for the mean $f_{\mathrm{spot}}$ value for each population. These binaries are removed in the subsequent analysis in the Pleiades.}
\end{figure}

With our spectroscopic $f_{\mathrm{spot}}$ and $T_{\mathrm{eff}}$ measurements, defined by the model of ambient and spot components on a stellar photosphere, we performed a self-consistent lookup on the SPOTS grids for the modified convective turnover timescale \citep{2020ApJ...891...29S}. This procedure yielded a self-consistently starspot-modelled Rossby number, which we discussed in detail in Section \ref{sec:rossbystarspots}. The relationship between the Rossby number and the starspot filling fraction in the Pleiades is demonstrated in Figure \ref{fig:SP2aI_fspot_Ro_binaries_big}.

To illustrate the importance of binary rejection, the binaries which are rejected from subsequent analyses are overplotted on a period-starspot filling fraction diagram in Figure \ref{fig:SP2aI_fspot_Ro_binaries_big}. The systematic offset between the identified binary population from our careful binary rejection technique in Section \ref{sec:binaryrej} is clearly visible in the rotationally saturated domain. These saturated rapid rotators are disproportionately low mass stars, such as late-K and M dwarfs; we note that the presence of an unresolved low-mass companion leads to a systematically higher starspot filling fraction for these stars than for the higher mass stars which are predominantly unsaturated in activity, as the relative flux contribution from the secondary is much higher when the primary is also a low mass star.

We visually demonstrate our starspot measurement process and analyze the concordance of our technique with the literature for a weak-lined T Tauri star in Taurus (Section \ref{sec:analysislkca4}) and a rapid rotator in the Pleiades (Section \ref{sec:analysishii296}).

\subsubsection{Starspots on a Weak-Lined T Tauri star: LkCa 4}\label{sec:analysislkca4}
The young star LkCa 4 is a pre-main sequence star in the Taurus star forming region with an age of $\sim$ $0.9$ Myr \citep{2015MNRAS.448..484L}. Weak-lined T Tauri stars are known to have strong kilogauss-strength magnetic fields \citep{2014MNRAS.444.3220D, 2018MNRAS.480.1754N, 2021MNRAS.504.2461N, 2007ApJ...664..975J} without an indication of disks or ongoing accretion \citep{2004Ap&SS.292..619V, 2014MNRAS.437.3202J}. LkCa 4 is a weak-lined T Tauri star with a detected mass accretion rate of $< 1.9 \times 10^{-9}$ $M_\odot / \mathrm{yr}$ \citep{2001ApJ...556..265W}. LkCa 4 in particular is well-studied, being observed with ZDI to have a predominantly dipolar poloidal $\sim$2 kG field and a toroidal $\sim$1 kG field \citep{2014MNRAS.444.3220D}. \citet{2017ApJ...836..200G} applied a spectral fit with a synthetic two-temperature model to the echelle spectra from the IGRINS instrument, obtaining large starspot filling fractions from different spectral orders. The best fit value expressed by \citet{2017ApJ...836..200G} by fitting to all spectral orders was $f_{\mathrm{spot}} = 80\% \pm 5\%$, with a cool component at $2750$ K and a hot component at $4100$ K.

\begin{figure}
\includegraphics[trim={0cm 0cm 0cm 0cm},width=\columnwidth]{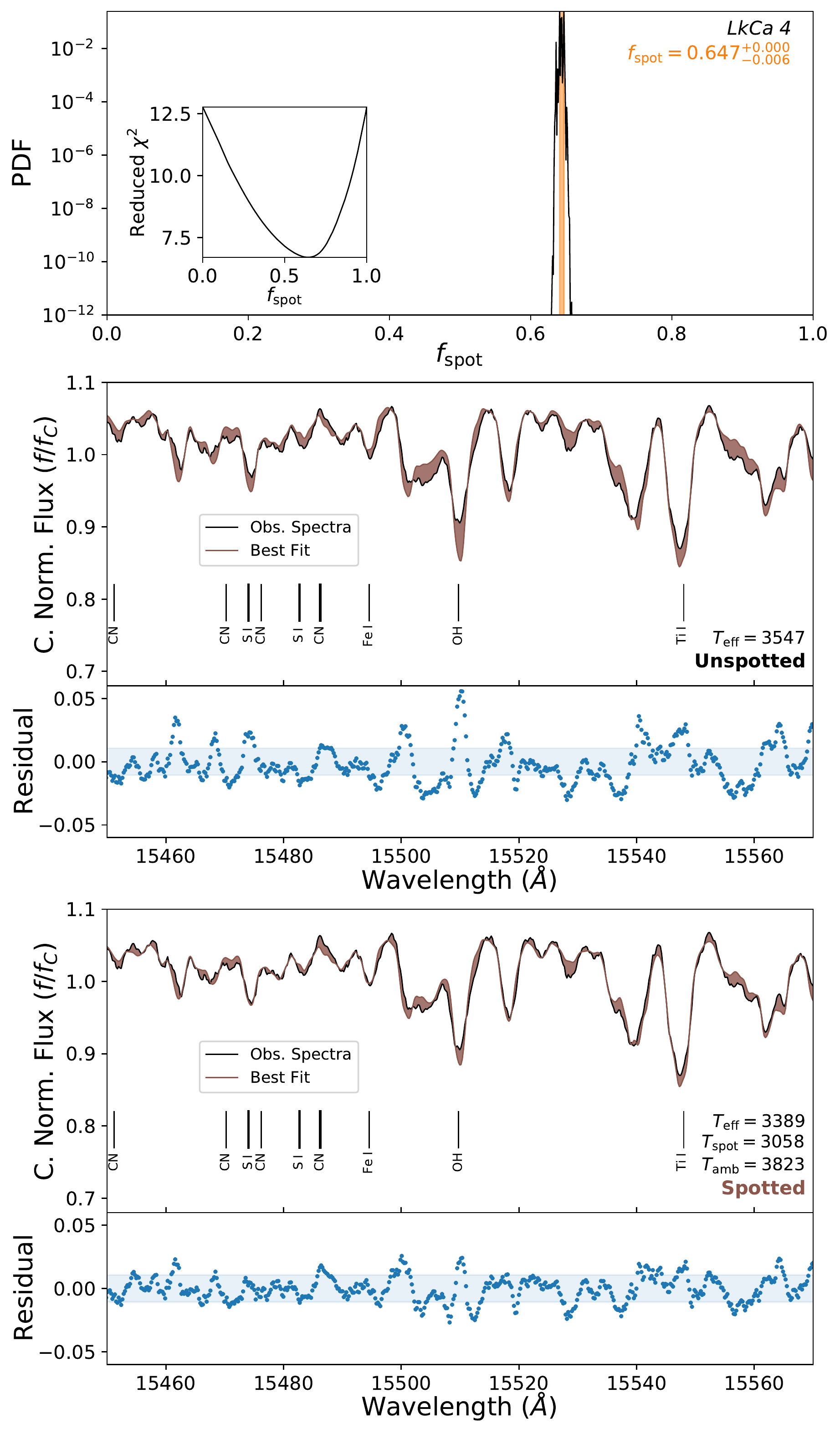}
\vspace{-1.5em}
\caption{\label{fig:SP2aI_LkCa4_big} Top: The probability density function of $f_{\mathrm{spot}}$ for LkCa4 and the reduced $\chi ^2$ surface. Middle: The observed and best fit continuum normalized spectrum with a single-temperature model along with a residual plot; the area in between the model and observation is filled in. Bottom: The observed and best fit continuum normalized spectra with the area in between filled in as before, but for the best fit two-temperature model. Visually it is apparent that the continuum normalized fluxes become much closer to the observation with the two-temperature model; statistically it is apparent from looking at the residual plots. Lines plotted are from the \textit{apogee} tool \citep{2016ApJ...817...49B}.}
\end{figure}

In Figure \ref{fig:SP2aI_LkCa4_big} we show our spectral fit to LkCa 4. We find a best fit $f_{\mathrm{spot}} = 64.7\%$ with a $1 \sigma$ confidence interval between $64.7\%$–$65.3\%$. Systematic errors can be significant for spectroscopic properties; we therefore compare our results to those inferred from IGRINS spectra. \citet{2017ApJ...836..200G} reports results for individual spectral orders, which are windows that are a little more than $\sim$200{\AA} wide. A comparison for spectral orders which overlap with the APOGEE H-band spectra include IGRINS orders $107$–$118$. Within these orders the starspot measurement ranges from a low of $38_{-36}^{+27} \%$ for order 107 to a high of $83_{-17}^{+6}\%$ in order 110 \citep{2017ApJ...836..200G}. Many of these orders include answers consistent with our best-fit value of $64.7\%$. To quantify this, out of 46 spectral orders ranging from 1.45–2.45 $\mu$m which produced estimates in the IGRINS spectra, our answer of $\sim$ $65\%$ lies within the $\pm 1 \sigma$ bound of the spectral order 28 times ($61\%$ of the spectral orders). Even though the IGRINS fit to the whole spectrum across 1.45–2.45 $\mu$m prefers a higher value of $f_{\mathrm{spot}}$, individual spectral orders seem to be distributed roughly as expected if our best fit value were accurate. We note that \citet{2017ApJ...836..200G} uses theoretical linelists and we use the APOGEE empirical linelists \citep{2021AJ....161..254S}---this may be a source of systematic between the two techniques, and in the domain of high reduced $\chi ^2$, may contribute to difficulties in obtaining fit constraints. We therefore conclude that our results are consistent with this literature comparison.

\subsubsection{Starspots on a magnetic rapid rotator: HII 296}\label{sec:analysishii296}
The Pleiad HII 296 (V966 Tau) has been extensively studied in the literature as a late-G or early-K type star which is magnetic, active, and rapidly rotating. In our sample, this star is Rossby-saturated with $\log \mathrm{Ro} = -0.70$, and not flagged as anomalously spotted. \citet{2016MNRAS.457..580F} inferred magnetic field properties for this star using Zeeman Doppler Imaging, inferring a mean surface magnetic field of $80.4$ G with a peak value of $273.6$ G, which is primarily poloidal ($89.5\%$ compared to $10.5\%$ toroidal) and dipolar ($62.7\%$ of the poloidal component, the rest in higher order fields).
To compare our starspot inference with the ZDI measurement, we assume that the mean surface field and peak field strength measured by ZDI are related by a factor of $f_{\mathrm{spot}}$. This would be the case in the limit where the magnetic field originates only in regions which are spotted. This is most likely an underestimate, since the mean surface field measured by ZDI is not sensitive to the B-field contribution from small-scale magnetic fields while the peak field is likely to be unaffected. With these assumptions, the inferred filling fraction from ZDI is approximately $29.3\%$. We prefer the comparison of this filling fraction proxy to an absolute comparison of magnetic field strengths since there is a systematic between magnetic measurements which we document in Section \ref{sec:magneticfield}; if the systematic between methods is dominated by a scale factor difference, the filling fraction should be comparatively less affected by systematics.

\begin{figure}
\includegraphics[trim={0cm 0cm 0cm 0cm},width=\columnwidth]{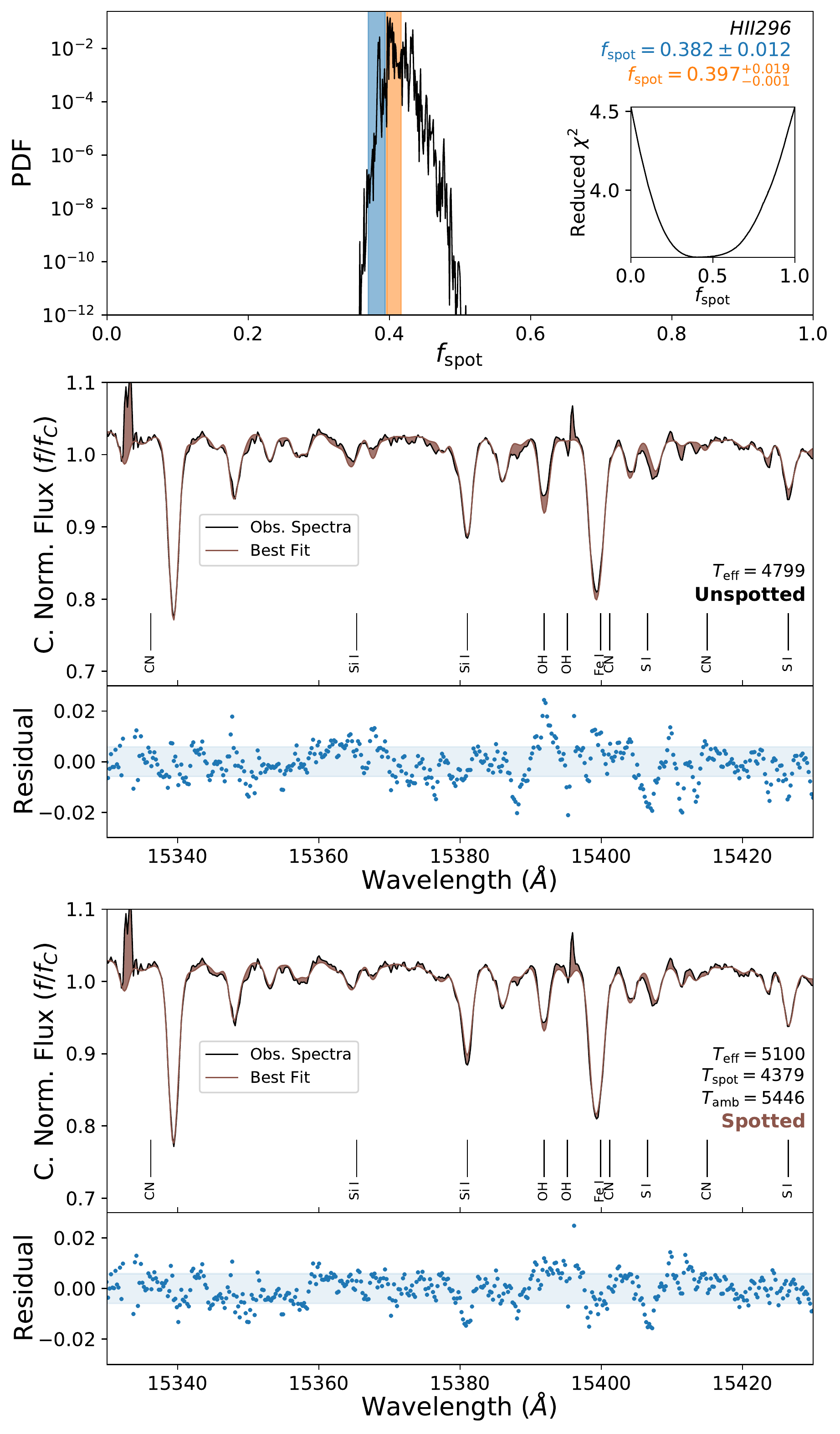}
\vspace{-1.5em}
\caption{\label{fig:SP2aI_HII296_big} Top: The probability density function of $f_{\mathrm{spot}}$ for HII 296 and the reduced $\chi ^2$ surface. For the middle and bottom plots, the same convention is used as in Figure \ref{fig:SP2aI_LkCa4_big}. A much broader distribution is seen in the reduced $\chi ^2$ surface and the improvement in fit with the continuum normalized flux is not as dramatic as with the weak-lined T Tauri star LkCa 4, but improvements in lines such as Si I are seen. Lines plotted are from the \textit{apogee} tool \citep{2016ApJ...817...49B}.}
\end{figure}

We see our starspot estimates for this star in Figure \ref{fig:SP2aI_HII296_big}; the blue band corresponding to the estimate from the UOBYQA algorithm (Section \ref{sec:gridconstruction}), and the orange band corresponding to an analysis of the $\chi ^2$ curve. In this case the two solutions differ by a systematic of $1.5\%$ on a star with a $39.7\%$ starspot coverage, slightly larger than the $1$$\sigma$ error estimate. However, a generic feature of these Pleiads is that they have a broader bottom to their $\chi ^2$ distributions than for LkCa 4, with a larger range of potential starspot filling fractions which are appropriate fits for the spectrum. A number of possibilities may explain this, including incomplete linelists, starspot structure, and reduced sensitivity of lines in the H-band for these stars; however, a dedicated analysis is necessary to characterize this feature.

Compared to our estimate of $39.7^{+1.9}_{-0.1} \%$, it is clear that the ZDI estimate of $29.3\%$ is smaller. It is possible that this is a systematic between the ZDI and two-temperature methods; however, cyclic spot variation, or systematics in the lines and spectral range may also account for the bias. It is remarkable that the detection of two temperature components on the spectrum of this star provides a reasonably similar signal to the complementary ZDI method, which is sensitive to spot modulation instead. Continued comparisons between the ZDI and two-temperature techniques will help characterize the relationship between these magnetic measurements.

\subsection{Pleiades Starspot Correlations}\label{sec:spotcorrel}
We compute Rossby numbers for our Pleiades targets using SPOTS convective overturn timescales (Section \ref{sec:rossbystarspots}) and rotation periods from \citet{2016AJ....152..113R}. In the single-star sequence identified in the Pleiades the bounds in log Ro are roughly from -2 to -0.3. For reference the log Rossby number of the Sun is $0.39$. We will use the starspot convective overturn timescale to infer Rossby numbers for all of our activity diagnostics.

\subsubsection{Starspot---Rossby Relation}\label{sec:spotrossby}
Our Pleiades starspot filling fractions as a function of Rossby number are displayed in Figure \ref{fig:SP2aI_fspot_Ro_binaries_big}. The overall pattern is similar to that for traditional activity diagnostics: a steep rise in filling fraction at large Rossby number followed by a flattening for low Rossby number. There may be signs of supersaturation in the most rapid rotators---a decline in coronal activity proxies with increasing rotation rate for Rossby-saturated stars \citep{1996A&A...305..785R,2011MNRAS.411.2099J,2016A&A...589A.113A,2022ApJ...931...45N}; due to the small number of sample stars rotating rapidly enough to probe supersaturation, we leave this discussion for future work. Our technique measures the mean surface filling fraction for spots, and therefore the mean surface magnetic field strength. Activity proxies for chromospheric and coronal heating could saturate for other reasons (for example, changes in field topology or heating mechanisms). Our data confirms that the saturation phenomenon is directly tied to stars reaching a maximum average field strength, which is consistent with recent direct magnetic measurements \citep{2022A&A...662A..41R, 2014MNRAS.441.2361V}. However, there remains a high degree of scatter at low Rossby, which we quantify as follows.

First, we remove all binaries from the sample. There are then two sources of scatter in the measured starspot filling fractions: the instantaneous measurement uncertainty and variations in the true spot filling fraction over time from stochastic effects and starspot cycles. We fit the mean starspot filling fraction---Rossby number relationship with a maximum likelihood estimator using 32 MCMC walkers in {\sc{emcee}} \citep{2013PASP..125..306F}, running for a length of $1.2 \times 10^4$ steps with a burn-in of $2500$ steps. The numerical fits took as inputs the logarithm of the Rossby number, the inferred starspot filling fraction, and the error in that starspot filling fraction for all single stars in the Pleiades; these were the points in x, y, and y errors respectively. These were then fit using a maximum likelihood estimator which had as its four free parameters: the critical x-coordinate at saturation, the saturation level as a y-coordinate, the slope of the function in the unsaturated domain, and an error inflation parameter---representing that the observed scatter in the data is too large to be explained by the random errors alone.

We use this error inflation parameter, which we name ``cycle variation'', to more accurately determine the uncertainties in each of our fit parameters; without accounting for this additional source of error, the uncertainties in each best-fit model parameter would be unphysically small, due to the small random error in each measurement. To motivate the form of this cyclic variation, we analyzed the underlying distribution of $f_{\mathrm{spot}}$ values in the saturated domain. We found that the distribution of starspot filling fractions for saturated stars is non-Gaussian, as the tails of a normal distribution are too long for the observed $f_{\mathrm{spot}}$ distribution. This was verified quantitatively using Kolmogorov-Smirnov tests; these showed that all sums of any normal distribution with the normally distributed random error were discrepant from the data with a p-value below 5\%. We instead modelled the data using a sum of a Gaussian component (its width set by the random errors in the measurements) and a uniform component, whose value was to be fit with the maximum likelihood estimation. This additional uniform component was found to reproduce the distribution very well, producing distributions indistinguishable from the data at the 5\% level between values of $0.51 < \log f < 1.19$, where $y_{\mathrm{err}} \times f$ represents the characteristic size of the uniform distribution.

The non-Gaussianity in the scatter of the saturated starspot measurements suggests that our measurements are significantly more precise than the non-Gaussian cyclic variation. If inflated random errors from the starspot measurements were the dominant noise source for the saturated stars, we would expect the variation to be Gaussian. However, there are a couple of caveats to this interpretation; first, it may well be that stellar variability cycles may not be distributed uniformly, but may include preferred high and low states better described by a bimodal distribution. Secondly, perhaps cycle variation is not the dominant parameter which changes the locations of stars on the activity---Rossby diagram. Time-domain studies in stellar activity and magnetism will resolve these and other questions.

Since the distribution is the sum of a uniform component set by the cyclic variation parameter and a Gaussian component informed by the random error, the likelihood function of the resultant distribution is the convolution of the two distributions:

\begin{equation}\label{eqn:likelihoodconvol}
p(\hat{y}) = \frac{1}{f y_{\mathrm{err}}} \exp \left[ \erf \left( -\frac{f}{\sqrt{2}} - \frac{(\hat{y} - y)}{\sqrt{2} y_{\mathrm{err}}} \right) - \erf \left( \frac{f}{\sqrt{2}} - \frac{(\hat{y} - y)}{\sqrt{2} y_{\mathrm{err}}} \right) \right] .
\end{equation}

Where $\hat{y}$ is the model with the saturation parameters as inputs, and $\erf$ the error function. Since the difference between error functions can be subject to rounding errors at extreme values, we compute this difference using the complementary error function $\erfc$ with the correct sign to avoid underflow. To prevent the cycle variation parameter $\log f$ from growing unphysically large, we limit its permissible values to the range of $\log f$ which cannot be distinguished from data at the 5\% level in the earlier K-S test---this means that the cycle variation parameter remains consistent with the data.

\begin{figure}
\includegraphics[trim={0cm 0cm 0cm 0cm},width=\columnwidth]{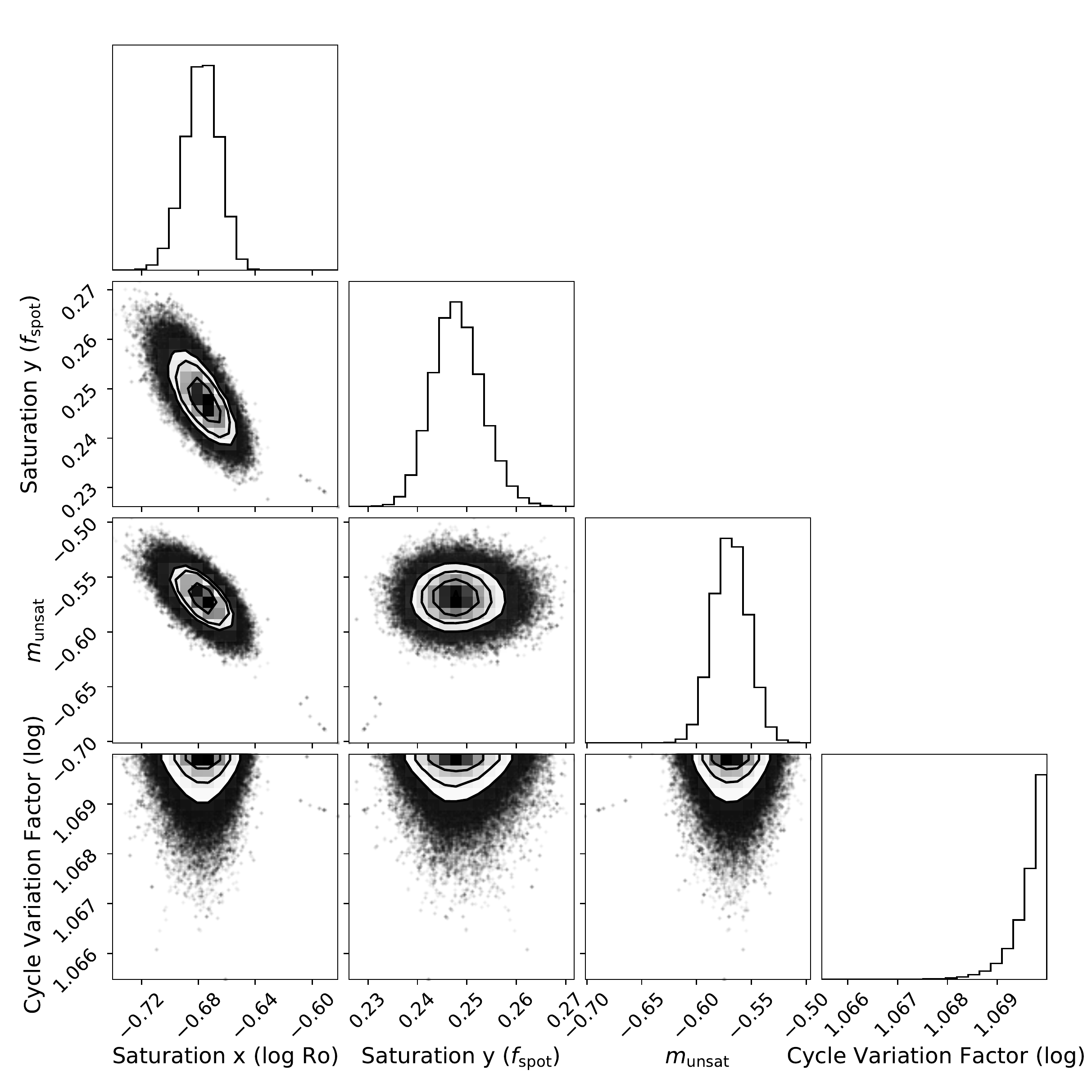}
\vspace{-1.5em}
\caption{\label{fig:SP2aI_ple_mcmc_big} MCMC parameters for the maximum-likelihood fit on the Pleiades single star sequence. The three recovered parameters correspond to the logarithm of the Rossby number at saturation (log $\mathrm{Ro}_{\mathrm{sat}} = -0.677_{-0.012}^{+0.012}$), the saturation level in starspot filling fraction ($f_{\mathrm{spot,}\, \mathrm{sat}} = 0.248_{-0.005}^{+0.005}$), and the slope of the unsaturated domain ($m_{\mathrm{unsat}} = -0.569_{-0.016}^{+0.016}$).}
\end{figure}

The MCMC chain in Figure \ref{fig:SP2aI_ple_mcmc_big} is well-sampled and well-converged with the expected degeneracy between the truncation point and the slope of the unsaturated domain. The converged free parameters then lead to the following logarithmic best fit relation:

\begin{equation} \label{eqn:spotrossby}
f_{\mathrm{spot}} = 
    \begin{cases}
      0.248, & \text{for}\ \mathrm{log \; Ro} < -0.677 \\
      -0.569 \; \mathrm{log \; Ro}-0.137, & \text{for}\ \mathrm{log \; Ro} \geq -0.677
    \end{cases} ,
\end{equation}

Performing this analysis again with a linear Rossby number as the x-coordinate and excluding the flagged anomalously active stars leads to the power law best fit relation:

\begin{equation} \label{eqn:spotrossby_pwr_law}
f_{\mathrm{spot}} = 
    \begin{cases}
      0.278, & \text{for}\ \mathrm{log \; Ro} < -0.729 \\
      6.34 \times 10^{-2} \; \mathrm{Ro}^{-0.881} , & \text{for}\ \mathrm{log \; Ro} \geq -0.729
    \end{cases} .
\end{equation}

\begin{figure*}
\includegraphics[trim={0cm 0cm 0cm 0cm},width=\textwidth]{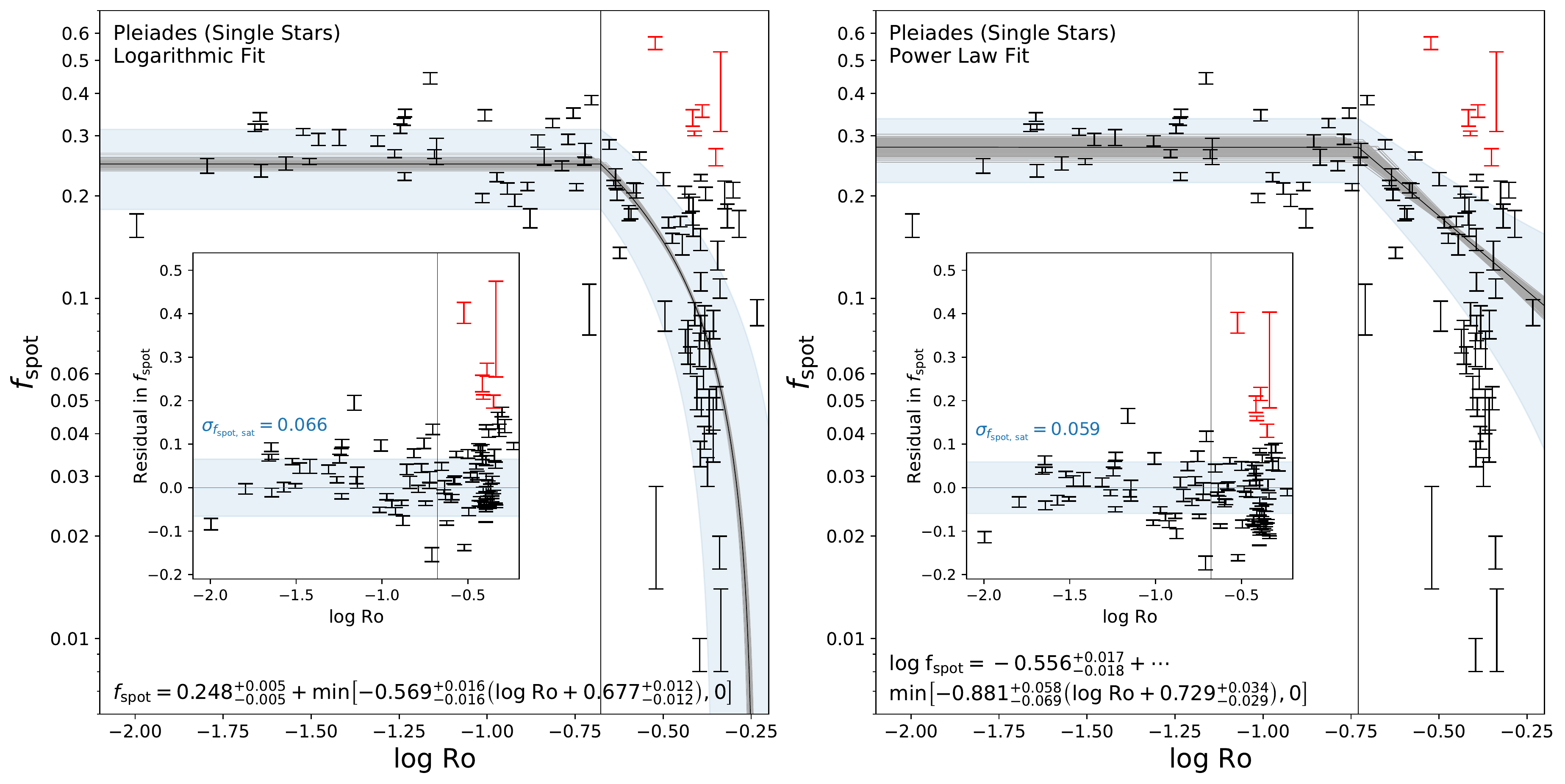}
\vspace{-1.5em}
\caption{\label{fig:SP2aI_ple_rossby_final_tp_big} Spot---Rossby Diagram with a two-component MCMC fit, including the uniform parameter for cycle variation. Left: logarithmic fit to the starspot filling fraction as a function of Rossby number. Right: power law fit to the same. The inset plots represent the residual of measured and best-fit starspot filling fractions against Rossby number. The light blue band represents the mean dispersion in spot measurements about the trend line in the saturated (log Ro $< -0.677$) regime. Red points are stars identified as more than $\pm 3 \sigma$ discrepant from the mean using the dispersion identified from the saturated domain from the left plot; the right plot excludes these points from the fit.}
\end{figure*}

These fit are then shown against the single-star Rossby sequence in Figure \ref{fig:SP2aI_ple_rossby_final_tp_big} as the left and right panels respectively. With both fits, the single-star activity relationships appear to show a saturation in starspots at rapid rotation, with a fall-off in the unsaturated domain. Below the saturation point identified by MCMC the dispersion in $f_{\mathrm{spot}}$ for saturated stars with the logarithmic fit is $\sigma_{f_{\mathrm{spot}}} = 0.066$; for the power law fit, the comparable dispersion is $\sigma_{f_{\mathrm{spot}}} = 0.059$. We use this saturated dispersion as the overplotted blue $1 \sigma$ band around the best-fit relation given by Equation \ref{eqn:spotrossby} \& \ref{eqn:spotrossby_pwr_law} respectively. There appears to be a small complement of stars which appear to be anomalously spotted for their slow rotation, identified strictly by a statistical criterion. These anomalously overspotted stars are highlighted red in the top-right panel of Figure \ref{fig:SP2aI_CMDs_big}; as shown, they pass each of our binary rejection cuts and appear to be on the single star sequence.
These stars are flagged and represented in red in the following Pleiades activity plots.

\subsubsection{Equipartition Magnetic Field---Rossby Relation}\label{sec:magneticfield}

Starspot filling fraction measurements can be used to estimate the mean equipartition magnetic field strength on the stellar surface. The relationship between the surface pressure and the equipartition field strength relies on a pressure balance argument:

\begin{equation}
\frac{B^2 _{\mathrm{eqp, \; max}}}{8 \pi} = \frac{P_{\mathrm{surf}}}{2} ,
\end{equation}

Where the equipartition magnetic field is the maximal field strength within an starspot to be in equipartition with the stellar photosphere. With an interpolation on the SPOTS \citep{2020ApJ...891...29S} isochrones and a measured $f_{\mathrm{spot}}$ and $T_{\mathrm{eff}}$, we derive an estimate for the stellar surface pressure. Then, the surface mean field is defined as the parameter:
\begin{equation}\label{eqn:beqpm}
\left< \left| B _{\mathrm{eqp}} \right| \right> = f_{\mathrm{spot}} \left| B _{\mathrm{eqp, \; max}} \right| .
\end{equation}

\begin{figure}
\includegraphics[trim={0cm 0cm 0cm 0cm},width=\columnwidth]{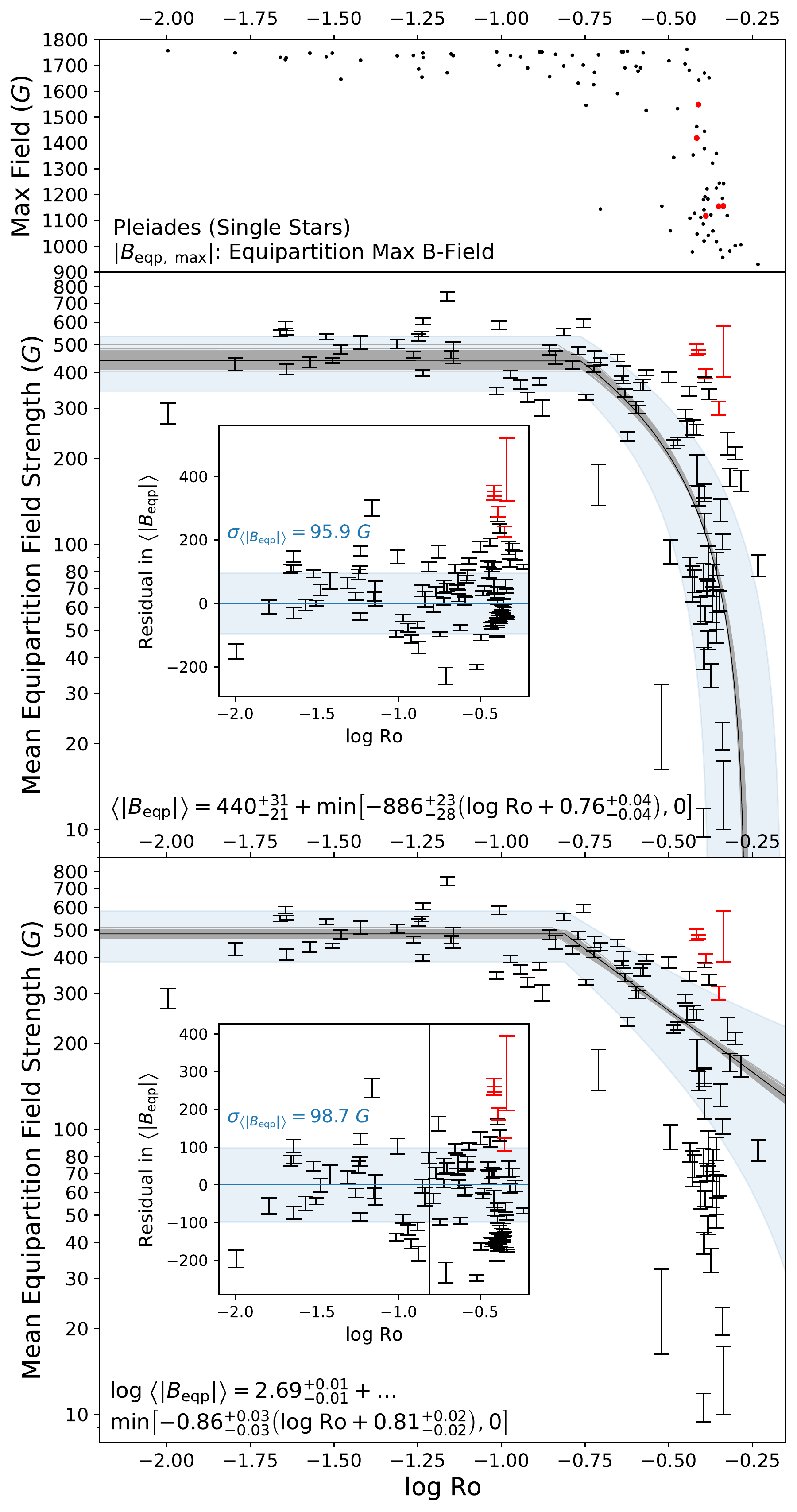}
\vspace{-1.5em}
\caption{\label{fig:SP2aI_BField_sud_big} Equipartition magnetic field strengths for Pleiads inferred from starspot filling fractions and SPOTS \citep{2020ApJ...891...29S} isochrones. Top: maximum field strength supported under equipartition; middle and bottom: a logarithmic and power law fit to the mean equipartition field strengths, respectively. Red indicates anomalously spotted single stars identified in Section \ref{sec:spotrossby}, and black indicates the rest of the single stars.}
\end{figure}

We demonstrate these magnetic field measurements in Figure \ref{fig:SP2aI_BField_sud_big}, with maximum likelihood fits similar to those used in Section \ref{sec:spotcorrel}; the equivalent K-S interval at the 5\% level we found here was $0.51 < \log f < 1.13$. Accounting for this non-Gaussianity in the cyclic variation parameter, we find the logarithmic and power law fits in the middle and bottom plots. The surface mean magnetic field strength saturates at a lower Rossby number than starspots, at a $\sim$$2 \sigma$ level in each fit. This lower saturation threshold---even using the same Rossby number definition---appears to be robust in this work, but needs to be studied systematically with more complete samples in order to assign it physical significance. For instance, the discrepancy between the starspot and magnetic field saturation thresholds may be a feature of equipartition magnetic fields weakening for higher mass stars at the same age, but may also reflect the mass-dependent saturation trend of stars in the Pleiades.

\begin{figure}
\includegraphics[trim={0cm 0cm 0cm 0cm},width=\columnwidth]{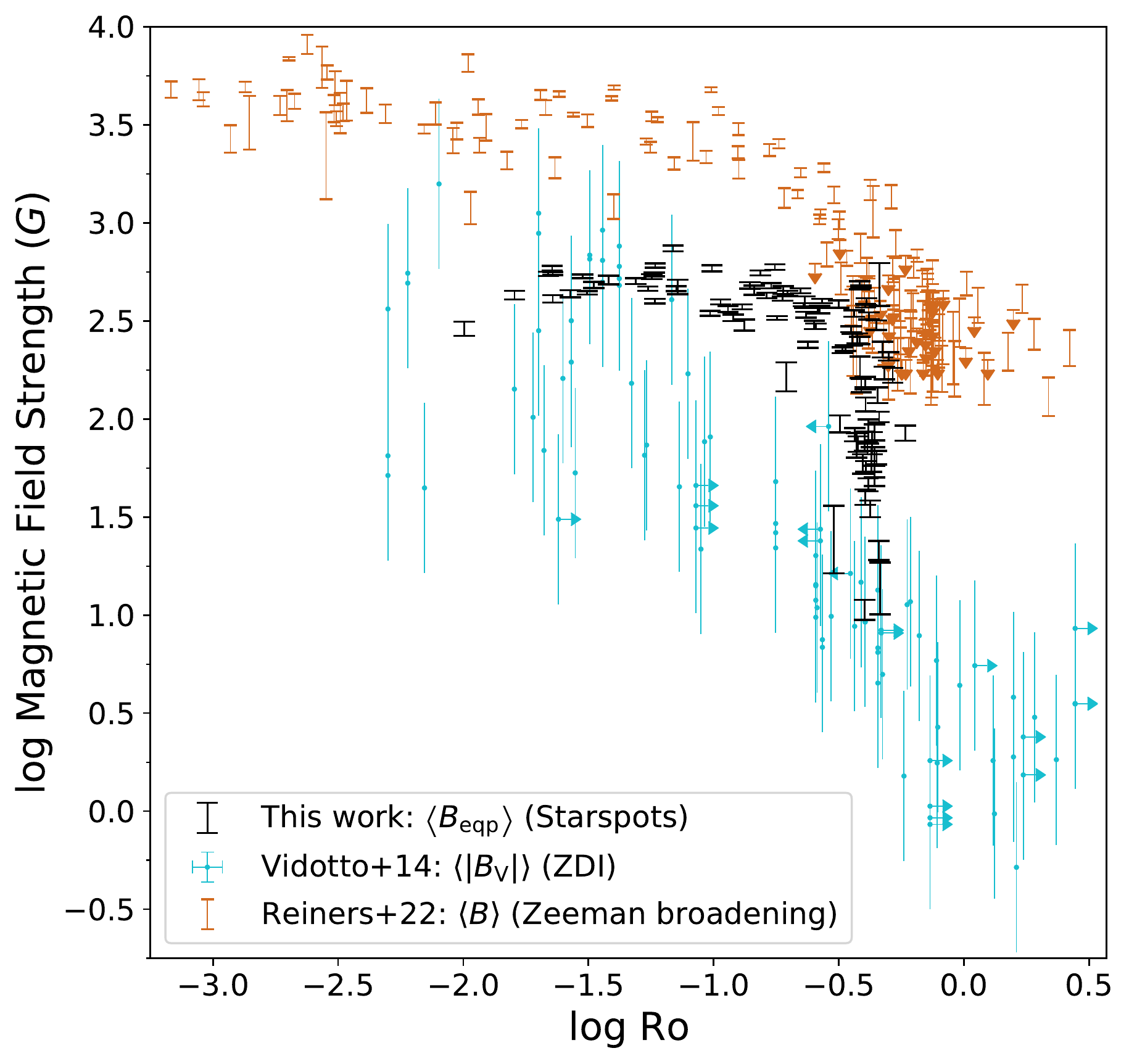}
\vspace{-1.5em}
\caption{\label{fig:SP2aI_BField_comparison_uns_big} A comparison of our equipartition magnetic field strengths for the Pleiades and magnetic measurements in the literature for field populations using ZDI \citep{2014MNRAS.441.2361V} and Zeeman Broadening \citep{2022A&A...662A..41R}. Errors are as presented in each work respectively, including the conservative $0.434$ dex error on the ZDI measurements suggested by \citet{2014MNRAS.441.2361V}. Each type of magnetic field measurements appears to be on a different scale, with systematics between scales of $\sim$0.7 dex.}
\end{figure}

These equipartition magnetic field measurements provide a complementary technique to the methods of ZDI and Zeeman broadening. We provide a comparison between different samples of magnetic field measurements in Figure \ref{fig:SP2aI_BField_comparison_uns_big}. These measurements appear to be on different scales offset by $\sim$0.7 dex; however, as these measurements are made with distinct techniques on different samples, the origin of this systematic is unclear. There is a physical reason to suspect a scale factor offset between measurements---whereas ZDI probes signed large-scale fields and Zeeman broadening is sensitive to the total unsigned field, these starspot---equipartition magnetic fields are tied by definition to fields which are strong enough to partially suppress convection and support a cooler spot.
It may be that the scale factors found between the B-field measurements represent physical differences in the underlying stellar magnetic field probed. If this is true, comparisons between the different classes of magnetic field measurements may contain geometric information. We emphasize that an analysis of the systematics between magnetic field measurements is necessary before comparing absolute magnetic measurements between systems.

\subsubsection{Activity---Starspot Relations}\label{sec:spotproxy}
\begin{figure}
\includegraphics[trim={0cm 0cm 0cm 0cm},width=\columnwidth]{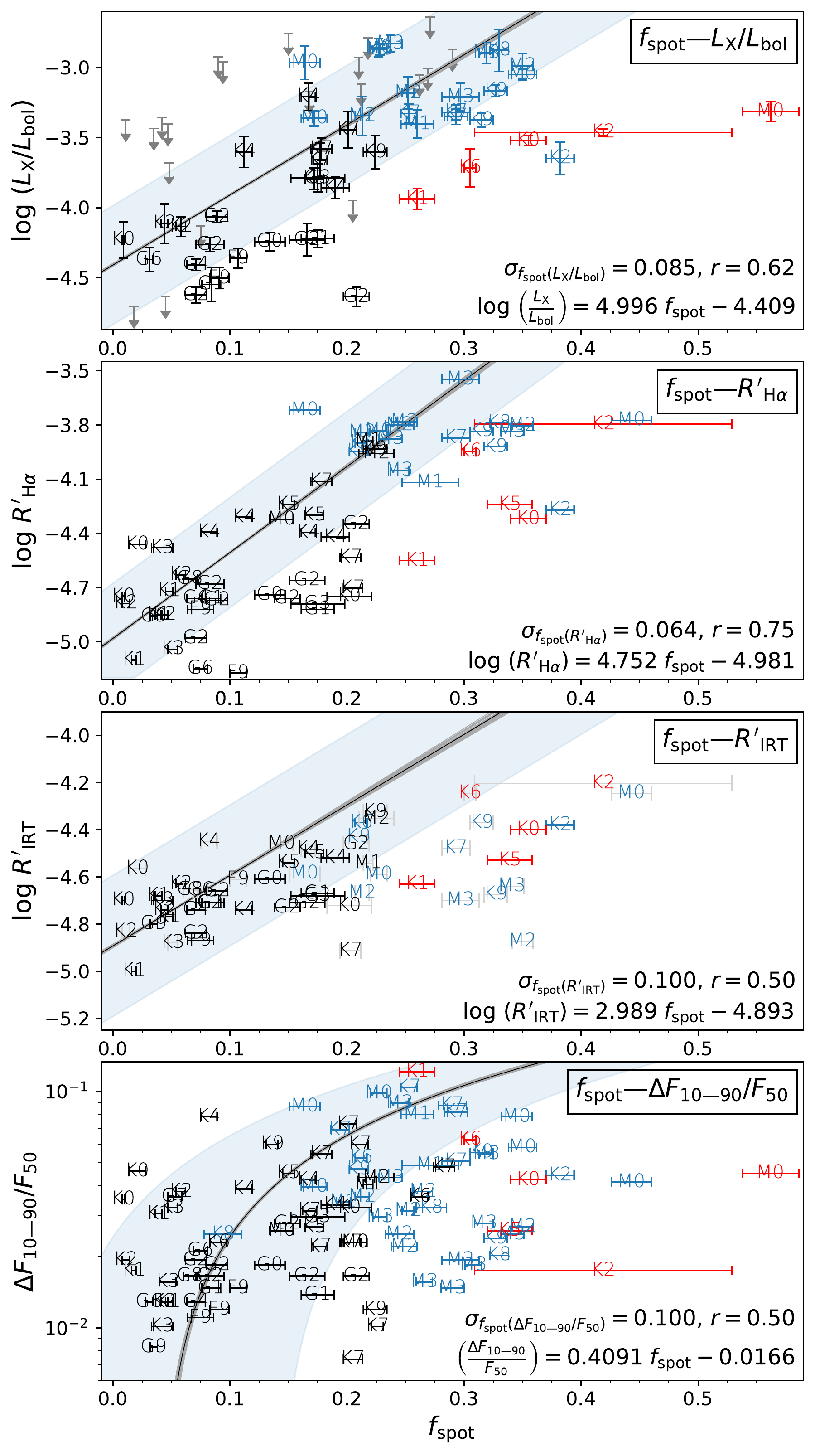}
\vspace{-1.5em}
\caption{\label{fig:SP2aI_Relations_big} Spot---Activity Proxy Diagram for our APOGEE Pleiades sample in X-rays (top), H$\alpha$ (second), Ca II IRT (third), and amplitude (bottom). Blue symbols indicate Rossby-saturated stars, black symbols indicate Rossby-unsaturated stars, and red symbols are the identified anomalously spotted stars; these selections are made using a cut at log Ro $= -0.677$ as derived from the maximum-likelihood fit to the starspot data (blue \& black) and the $\pm 3 \sigma$ outliers (red). Maximum-likelihood fits are computed for each plot and described in Section \ref{sec:spotproxy}. Top: a relationship between $f_{\mathrm{spot}}$ and log $\left( L_X / L_{\mathrm{bol}} \right)$ is observed with a dispersion of $\sigma_{f_{\mathrm{spot}}} = 0.085$. Second: a relationship between $f_{\mathrm{spot}}$ and log $R^{\prime}_{\mathrm{H}\alpha}$ with a dispersion of $\sigma_{f_{\mathrm{spot}}} = 0.064$. Third: a relationship between $f_{\mathrm{spot}}$ and log $R^{\prime}_{\mathrm{IRT}}$ with a dispersion of $\sigma_{f_{\mathrm{spot}}} = 0.100$; gray symbols indicate that the data were transformed from Ca H\&K measurements and should be interpreted with caution. Bottom: a relationship between $f_{\mathrm{spot}}$ and $\Delta F_{10-90}/F_{50}$ with a dispersion of $\sigma_{f_{\mathrm{spot}}} = 0.100$.}
\end{figure}

\begin{table}
 \caption{A description of the columns in the Pleiades online table.}
 \label{tab:ple}
 \begin{tabular*}{\columnwidth}{@{}l@{\hspace*{20pt}}l@{}}
  \hline
  Label & Contents \\
  \hline
  APOGEE ID & 2MASS ID for the source \\
  RA & Right ascension in decimal degrees (J2000) \\
  Dec & Declination in decimal degrees (J2000) \\
  Prot & Rotation period from \citet{2016AJ....152..113R} \\
  Ro & Theoretical Rossby number \citep{2020ApJ...891...29S} \\
  Bin\_Flag & Global binarity flag computed from Section \ref{sec:binaryrej} \\
  Bin\_Photometric & Photometric binary flag \\
  Bin\_Gaia\_RV & Gaia DR3 RV variance flag for binarity \\
  Bin\_APOGEE\_RV & APOGEE RV flag for binarity \\
  Bin\_Gaia\_RUWE & \verb|RUWE| flag for binarity \\
  Bin\_Gaia\_Multiple & Multiple source flag for binarity \\
  Teff & Two-temperature effective temperature [K] \\
  fspot & Two-temperature starspot filling fraction \\
  xspot & Two-temperature starspot temperature contrast \\
  logg & Two-temperature surface gravity \\
  {[}M/H{]} & Two-temperature metallicity \\
  vsini & Two-temperature rotational velocity [km/s] \\
  vdop & Two-temperature microturbulence [km/s] \\
  e\_Teff & Error in derived $T_{\mathrm{eff}}$ \\
  e\_fspot & Error in derived $f_{\mathrm{spot}}$ \\
  e\_xspot & Error in derived $x_{\mathrm{spot}}$ \\
  e\_logg & Error in derived logg \\
  e\_{[}M/H{]} & Error in derived {[}M/H{]} \\
  e\_vsini & Error in derived vsini \\
  e\_vdop & Error in derived microturbulence \\
  B\_eqp\_mean & Derived mean equipartition magnetic field [G] \\
  e\_B\_eqp\_mean & Error in mean equipartition magnetic field \\
  B\_eqp\_max & Derived max equipartition magnetic field [G] \\
  e\_B\_eqp\_max & Error in max equipartition magnetic field \\
  log\_Lx\_Lbol & Norm. fractional X-ray emission (Section \ref{sec:spotproxy}) \\
  e\_log\_Lx & Reported error in X-ray emission \\
  l\_log\_Lx & Upper limit flag in X-ray emission \\
  Ref\_log\_Lx\_Lbol & Reference for X-ray data \\
  log\_R'Ha & Excess chromospheric H$\alpha$ emission \\
  e\_log\_R'Ha & Reference for $R^{\prime}_{\mathrm{H}\alpha}$ \\
  log\_R'CaIRT & Excess chromospheric Ca II IRT emission \\
  e\_log\_R'CaIRT & Reference for $R^{\prime}_{\mathrm{Ca \; IRT}}$ \\
  frac\_flux\_10\_90 & Fractional flux difference, 10th to 90th percentile \\
  \hline
  \multicolumn{2}{l}{This table is available in its entirety in a machine-readable form.}\\ 
 \end{tabular*}
\end{table}

To see the correspondence of these new starspot detections with existing activity proxies, we collate a dataset consisting of X-ray, H$\alpha$, Ca IRT, and amplitude measurements in Table \ref{tab:ple}. First, the X-ray data is scaled to the same bolometric luminosity scale to remove systematics between different analyses. We first obtain a $BC_K$ for each star from its effective temperature using the \citet{2013ApJS..208....9P} online extended dwarf table\footnote{\url{https://www.pas.rochester.edu/~emamajek/EEM_dwarf_UBVIJHK_colors_Teff.txt}}; then, we obtain an absolute K-band magnitude using 2MASS photometry and {\gaia} DR3 parallaxes. Summing the two yields an estimate for $M_{\mathrm{bol}}$, which is converted to $L_{\mathrm{bol}}$ using as the solar bolometric magnitude and luminosity as a zero point. We then divide each $L_X$ by this $L_{\mathrm{bol}}$ to obtain these coronal emission data. The chromospheric data are transformed according to Appendix \ref{sec:appendixchromospheric} as the chromospheric excess measurements may have a different scale whether they are inferred with photometric excess or spectroscopic equivalent width excess; we scale everything onto the photometric $R^{\prime}$ scale where necessary using stars in common. For the Ca data, to augment the \citet{1993ApJS...85..315S} Pleiades measurements at lower masses we transformed from the measurements of the Ca K line from \citet{2018MNRAS.476..908F} to the infrared triplet; this is not an exact correspondence as they are separate lines and therefore the Ca plot should be interpreted with care. The amplitude data assessed in this section are provided also by \citet{2016AJ....152..113R} as the magnitude difference between the 10th and 90th percentile of the light curve. Since starspot filling fraction and flux variations across the star are physically related, we transform the amplitude in magnitudes to fractional flux variation of the star. By convention the amplitudes $\Delta m$ follow the relation:

\begin{equation}
F_{10}/F_{90} = 10^{-\Delta m/2.5} ,
\end{equation}

Then we have the subsequent relations:

\begin{equation}\label{eqn:F1090}
\Delta F_{10-90}/F_{90} = \frac{F_{90} - F_{10}}{F_{90}} = 1 - 10^{-\Delta m/2.5} ,
\end{equation}

To scale to the ``mean'' flux level in between the 10th and 90th percentile variations:

\begin{equation}\label{eqn:F50}
F_{50}/F_{90} = \frac{1}{2} \frac{ F_{10} + F_{90} }{F_{90}} = \frac{ 1 + 10^{-\Delta m/2.5} }{2} ,
\end{equation}

Dividing Equation \ref{eqn:F1090} by Equation \ref{eqn:F50}, this leads to an expression for the fractional flux variation scaled by the mean flux:

\begin{equation}
\Delta F_{10-90}/F_{50} = \left(1 - 10^{-\Delta m/2.5} \right) \left( \frac{2}{1 + 10^{-\Delta m/2.5}} \right) .
\end{equation}

We compare our activity proxies to starspot filling fraction in Figure \ref{fig:SP2aI_Relations_big}. There appears to be a correlation in all plots between starspot filling fraction and activity measurements with a sizable variation; this observed scatter may represent a flux variation for sources observed at different points in their activity cycles. The two activity proxies which correlate best with starspots are the H$\alpha$ and X-ray measurements; photometric amplitude and Ca II IRT have smaller correlation coefficients with $f_{\mathrm{spot}}$.

\subsubsection{Activity---Rossby Relations}\label{sec:rossbysat}

The next step is quantifying the detailed properties of different activity indicators. We are especially interested in two questions: whether different diagnostics saturate at the same level, and whether the stars identified as anomalous in the starspot---Rossby domain are unusual with other diagnostics. We perform a maximum likelihood fit with a modified likelihood function using a cycle variation parameter as described in Section \ref{sec:spotrossby}; the cycle variation parameter is limited again to the largest width of the uniform distribution which is indistinguishable from the saturated stars at the 5\% level. Only the X-ray data include upper limits, and we remove upper limits from our fits and subsequent analysis (see discussion in \citet{2011ApJ...743...48W}). By excluding upper limits in the X-rays, it is possible that the derived activity relation is biased below the true relation. However, we are using multiple sources of X-ray data which may have different methodologies for assigning upper limits, and this simplifies the subsequent analysis \citep{2011ApJ...743...48W}. For the chromospheric plots it is not a problem that errors are not provided since we use our cyclic variation parameter, which is set by the spread in the distribution.

\begin{figure}
\includegraphics[trim={0cm 0cm 0cm 0cm},width=\columnwidth]{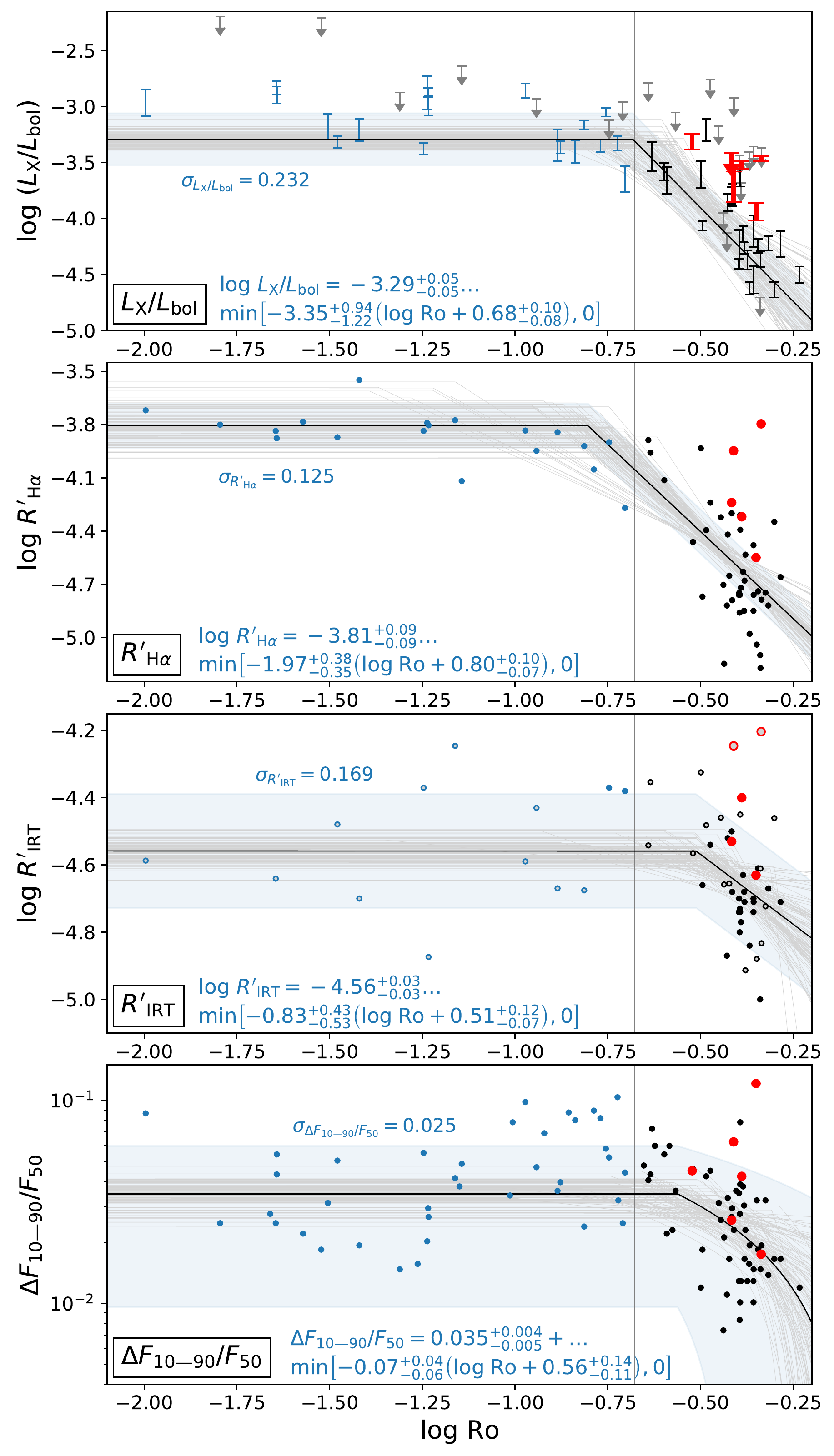}
\vspace{-1.5em}
\caption{\label{fig:SP2aI_Rofits_big} Rossby activity relations for our APOGEE Pleiades sample in X-ray (top), H$\alpha$ (second), Ca II IRT (third), and amplitude (bottom). Blue points indicate stars satisfying the saturated condition (log Ro $< -0.677$) from the fit to starspot filling fraction. Section \ref{sec:rossbysat} describes the procedure for obtaining the fit lines in black (which do not fit to upper limits or flagged anomalously active Pleiads by construction) and the light bands, which indicate the dispersion of the population. Black points the unsaturated condition (log Ro $> -0.677$) and red points indicate the identified $\pm 3 \sigma$ discrepant stars identified in Section \ref{sec:spotrossby}. In the Ca II IRT panel, points with unfilled centers indicate transformed LAMOST data using the method discussed in Appendix \ref{sec:appendixchromospheric} and should be interpreted with caution. The light gray vertical line indicates the saturation point found in APOGEE spectroscopic $f_{\mathrm{spot}}$ (log $\mathrm{Ro}_{\mathrm{sat}} = -0.677$).}
\end{figure}

The direct relationship between each activity proxy and Rossby number is shown in Figure \ref{fig:SP2aI_Rofits_big}, indicating Rossby-saturated stars in blue and Rossby-unsaturated stars in black according to whether they are saturated from the fit to $f_{\mathrm{spot}}$. Stars with unfilled symbols in the Ca IRT plot are the ones transformed from LAMOST Ca K data. The black fit lines for each activity proxy are obtained separately. We also include the anomalously active stars if they have X-ray, chromospheric, or amplitude data and plot them as red symbols. This allows a calculation of the z-score in activity for each anomalous star by dividing their departure from the best fit relation by the dispersion in Rossby-saturated stars, a calculation which is shown in Table \ref{tab:pleoss}. We see that the z-score of significance is high for nearly every activity proxy, indicating that these stars are offset from the mean relation.

As these measurements were non-simultaneous, the fact that these stars appear to be on the upper envelope or anomalously active at some significance in these activity proxies measurements is at odds with the idea that we have identified stars at the high point of their cyclic variation. If cyclic variation were the main cause of deviations from the trend line, we would catch most of these stars at significantly less active states in their cycle if the variation time is on the order of decades.

These departures from the standard Rossby relation are astrophysically interesting. There are a number of plausible mechanisms that could induce departures from a unique Rossby relationship---for example, core-envelope decoupling models predict large radial shears in young stars; fully convective stars could have a different dynamo mechanism than radiative core ones; or the structural effects of rotation could have mass-dependent feedback on the dynamo mechanism.

\begin{table}
 \caption{Activity z-scores for anomalously over-spotted single Pleiads.}
 \label{tab:pleoss}
  \begin{tabular*}{\columnwidth}{@{}l@{\hspace*{9pt}}l@{\hspace*{9pt}}l@{\hspace*{9pt}}l@{\hspace*{9pt}}l@{\hspace*{9pt}}l@{\hspace*{9pt}}l@{}}
  \hline
  & \multicolumn6c{Z-score} \\
  APOGEE ID & Spots & $\left< \left| B_{\mathrm{eqp}} \right| \right>$ & X-ray & H$\alpha$ & Ca IRT & Ampl \\
  \hline
2M03441307+2401509 & 5.56 & 4.39 &  4.24 &  7.42 & 2.96 &  -0.02 \\
2M03463889+2431132 & 3.65  & 3.62 &  $<$3.32 &  2.62 & 0.64 & 0.07 \\
2M03461174+2437203 & 3.02 & 2.35 & 2.01 & 1.17 & 0.37 &  4.08 \\
2M03463287+2318191 & 6.13 & ---  &  2.21 & --- & ---  & 0.53 \\
2M03473800+2328051 & 3.18 & 3.58 & 2.09 & 5.04 & 2.35 &  1.56 \\
2M03440484+2416318 & 4.13 & 3.00 & 3.25 & 2.41 & 1.54 &  0.81 \\
  \hline
 \end{tabular*}
\end{table}

It appears visually that the critical Rossby number at saturation changes with different activity indicators in Figure \ref{fig:SP2aI_Rofits_big}. \citet{2021ApJ...916...66B} found that the critical Rossby number differs across activity diagnostics, particularly between their mid-frequency continuum and starspot variability proxies.
This introduces the possibility that activity diagnostics sample distinct depths in the convection zone.
Since convective velocity and pressure scale height are both strong functions of depth in stars, phenomena anchored at a variety of scales may saturate at different Rossby thresholds.
As we include the cycle variation parameter in our fits, we can systematically study the saturation threshold in different activity proxies.

\begin{figure*}
\includegraphics[trim={0cm 0cm 0cm 0cm},width=\textwidth]{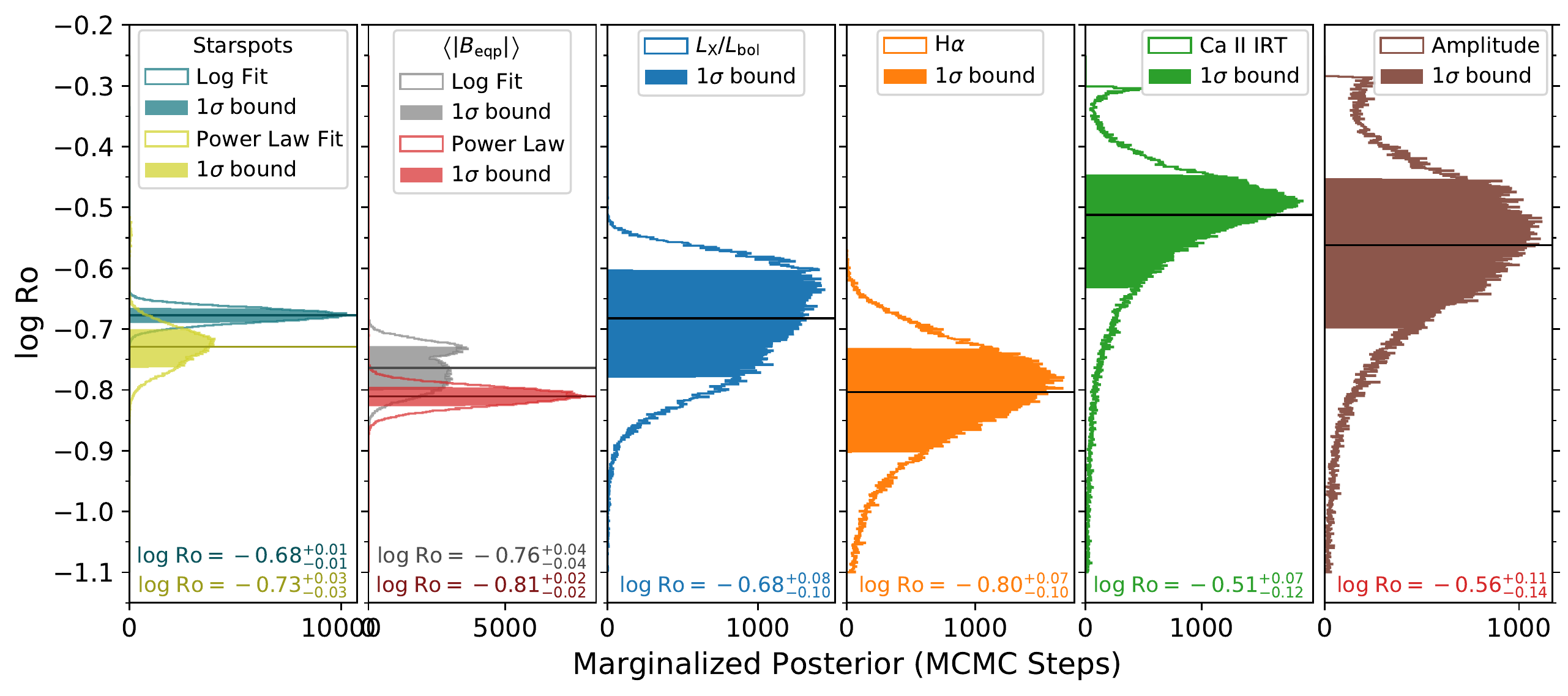}
\vspace{-1.5em}
\caption{\label{fig:SP2aI_ple_mcmc_logRo_loc_ps_big} Probability distributions of the critical Rossby number in our activity---Rossby fits, accounting for cycle variation in the posterior (Section \ref{sec:spotrossby}). From left to right, these are the critical Rossby numbers for our fits in starspots, equipartition magnetic field strengths, X-rays, H$\alpha$, Ca IRT, and amplitudes. We have plotted both the logarithmic and power law fits for the starspot and magnetic plots with different colors in each; the agreement is good between logarithmic and power law fits, but there is a $2\sigma$ offset between the threshold derived with starspots and magnetic measurements.}
\end{figure*}

We quantify the uncertainty in our threshold Rossby numbers by plotting the marginalized MCMC chains in Figure \ref{fig:SP2aI_ple_mcmc_logRo_loc_ps_big}. The uncertainty estimation is meaningful because the cycle variation parameter (Section \ref{sec:spotrossby}) in our modified likelihood function (Equation \ref{eqn:likelihoodconvol}) includes a uniform distribution which better reproduces these activity histograms in the saturated domain. Since we are accounting for the possibility that the distribution of points is not adequately described by the reported random errors, this allows us to examine the knee of the activity---Rossby fit. From left to right they show the starspot, equipartition magnetic field, X-ray, H$\alpha$, Ca IRT, and amplitude measurements, respectively. The colored band in each histogram represents the $1 \sigma$ uncertainty bound for that parameter. We find that the data are consistent with the same saturation break in all diagnostics, with no statistically significant difference in the critical Rossby number after accounting for possible cycle effects.

The Rossby number used for all of these calculations in Figure \ref{fig:SP2aI_ple_mcmc_logRo_loc_ps_big} are the ones from Section \ref{sec:rossbystarspots} accounting for the impact of starspots. For a discussion on the impact of non-spotted Rossby number definitions on the location of the transition between Rossby-saturated and unsaturated stars, see Appendix \ref{sec:appendixrossby}.
We also note that the width of the histograms in Figure \ref{fig:SP2aI_ple_mcmc_logRo_loc_ps_big} are minimized for the starspot and magnetic field measurements; this is consistent with a more precise measurement of the critical Rossby number. As the number of observations varies for each activity measurement, it is also possible that the Rossby number threshold will become more precise in other activity proxies as samples become larger.

\subsection{Starspots in M67}\label{sec:discm67}

\begin{table}
 \caption{A description of the columns in the M67 online table.}
 \label{tab:m67}
 \begin{tabular*}{\columnwidth}{@{}l@{\hspace*{20pt}}l@{}} 
  \hline
  Label & Contents \\
  \hline
  APOGEE ID & 2MASS ID for the source \\
  RA & Right ascension in decimal degrees (J2000) \\
  Dec & Declination in decimal degrees (J2000) \\
  Evstate & Evolutionary state as defined in Section \ref{sec:discm67} \\
  Bin\_Flag & Global binarity flag computed from Section \ref{sec:binaryrej} \\
  Bin\_Photometric & Photometric binary flag \\
  Bin\_Gaia\_RV & Gaia DR3 RV variance flag for binarity \\
  Bin\_APOGEE\_RV & APOGEE RV flag for binarity \\
  Bin\_Gaia\_RUWE & \verb|RUWE| flag for binarity \\
  Bin\_Geller17\_sl & Single-lined binary flag from \citet{2021AJ....161..190G} \\
  Bin\_Geller17\_dl & Double-lined binary flag from \citet{2021AJ....161..190G} \\
  Teff & Two-temperature effective temperature [K] \\
  fspot & Two-temperature starspot filling fraction \\
  xspot & Two-temperature starspot temperature contrast \\
  logg & Two-temperature surface gravity \\
  {[}M/H{]} & Two-temperature metallicity \\
  vsini & Two-temperature rotational velocity [km/s] \\
  vdop & Two-temperature microturbulence [km/s] \\
  e\_Teff & Error in derived $T_{\mathrm{eff}}$ \\
  e\_fspot & Error in derived $f_{\mathrm{spot}}$ \\
  e\_xspot & Error in derived $x_{\mathrm{spot}}$ \\
  e\_logg & Error in derived logg \\
  e\_{[}M/H{]} & Error in derived {[}M/H{]} \\
  e\_vsini & Error in derived vsini \\
  e\_vdop & Error in derived microturbulence \\
  \hline
  \multicolumn{2}{l}{This table is available in its entirety in a machine-readable form.}\\ 
 \end{tabular*}
\end{table}

In M67 the APOGEE H-band spectra available span a range of evolutionary states and binarity. With our sample we produce starspot measurements for M67 stars (Table \ref{tab:m67}). There are distinct domains of interest in this cluster. The turnoff stars are in the transition regime where starspots and activity appear to become weak from photometric proxies \citep{2021ApJS..255...17S}; evolved stars also have different activity properties than dwarfs. We therefore identify evolutionary states through logg and $T_{\mathrm{eff}}$ cuts and then display them on Figure \ref{fig:SP2aI_CMDs_big}.
The four stars with significant radial velocity variations above the subgiant branch have been flagged; these include two sub-subgiants identified by \citet{2017ApJ...840...66G}, S1063 and S1113.

Solar-type stars are relatively slow rotators and are therefore expected to be inactive---with the possible exception of synchronized binary stars. Turnoff stars in M67 are in the late F star domain, and have modest rotation rates typically around $5.4\pm2.4$ km/s, with a range in our sample between 1.6–11.8 km/s. We find that their expected Rossby numbers are likely large from a forward modelled exercise using \citet{2013ApJ...776...67V}; a decrease in Rossby number as these stars evolve to the subgiant phase is found in our exercise due to a lengthening convective turnover timescale \citep{2013ApJ...776...67V, 1985ApJ...299..286G}. From angular momentum conservation, as these stars ascend the giant branch, the single evolved stars are expected to become slow and inactive. Evolved blue stragglers (interacting binary systems) could very well appear as active evolved stars, however, and they are not rare in M67---reaching $\gtrsim$$8\%$ of the solar-type spectroscopic binary population \citep{2019ApJ...881...47L, 2021AJ....161..190G}. We therefore break our sample up into distinct regimes---solar-like, turnoff, subgiant, and giant---for our examination of starspots and rotation.

We removed binaries in the main sequence stars using the procedure from Section \ref{sec:binaryrej}, however it was impossible to use the photometric binary cut in the main sequence turnoff. Since the secondary contributes minimal light in the subgiant and giant branch, the photometric rejection also cannot be used during post-main sequence evolution. As a result, we primarily used the kinematic rejection techniques and the spectroscopic RV catalog from \citet{2021AJ....161..190G}. Photometric binaries in the turnoff may be systematically increasing the starspot signal in some of those stars, and it is possible that some number of them are interacting or merged systems. We also caution against the interpretation of these stars at higher temperature near $6500$ K because they may be rapidly rotating, which may lead to a false detection of starspots due to, for instance, limb- or gravity-darkening effects; alternatively, these detections may be real and indicate a change in geometry in the magnetic field for these hotter stars. Since it requires more careful analysis to determine whether two-temperature detections in this temperature regime ($\gtrsim 6500$ K) are real starspot signals, we resist interpreting our filling fractions in the M67 turnoff stars.

\begin{figure}
\includegraphics[trim={0cm 0cm 0cm 0cm},width=\columnwidth]{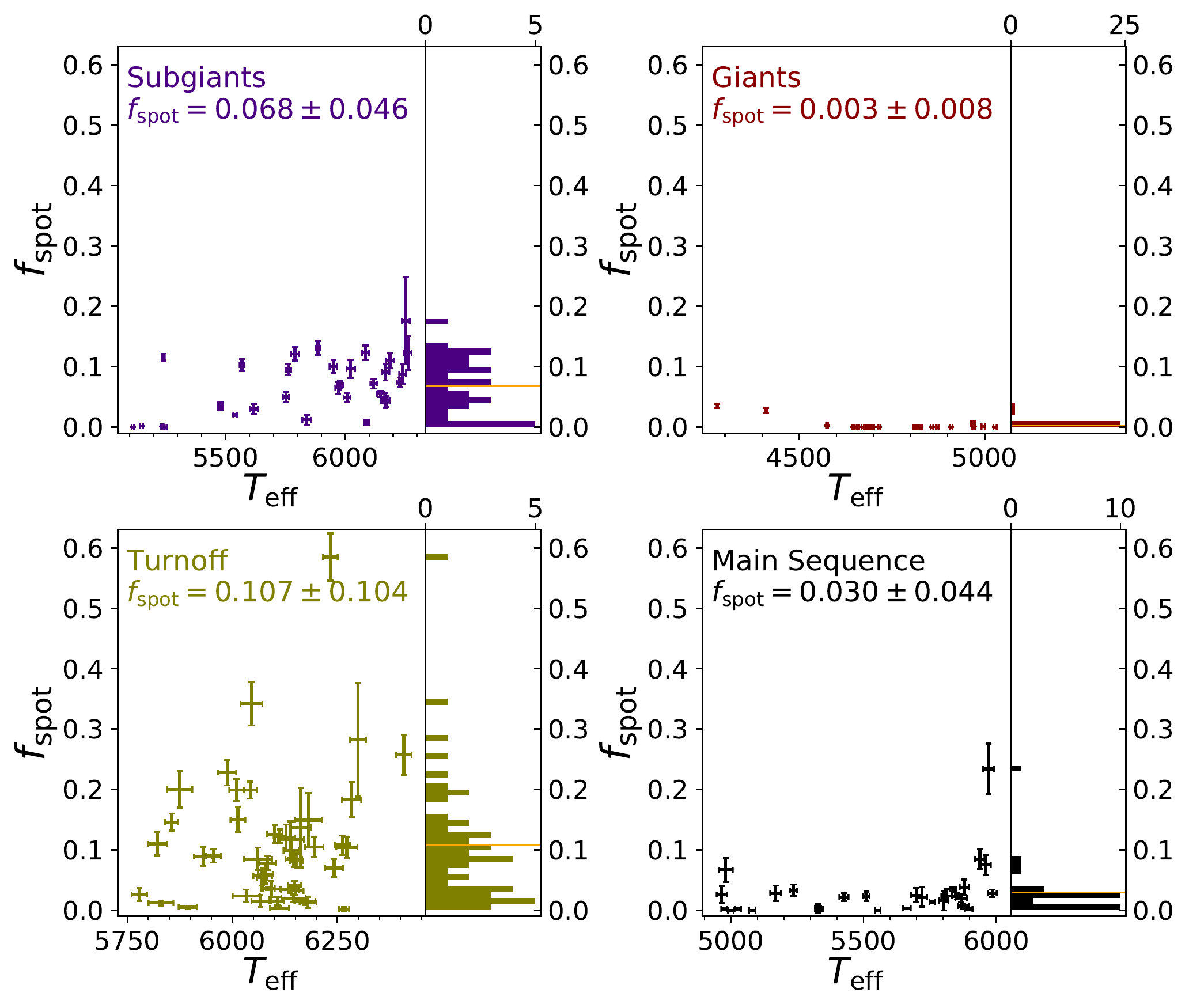}
\vspace{-1.5em}
\caption{\label{fig:SP2aI_M67_fspot_Teff_big} Starspot filling fractions for different evolutionary states in M67. From top left to bottom right, panels indicate stars flagged with the subgiant, giant, turnoff, and main sequence evolutionary states.}
\end{figure}

For each evolutionary state we can plot inferred starspot filling fraction against $T_{\mathrm{eff}}$ to identify statistical properties and trends. There is a low starspot filling fraction in giants and solar-type main sequence stars (Figure \ref{fig:SP2aI_M67_fspot_Teff_big}). The subgiants appear to be systematically higher in spot coverage than the main sequence stars, suggesting that there is a revival of activity prior to the ascent along the giant branch. \citet{1985ApJ...299..286G} predicted the existence of active subgiants with a lengthening convective turnover timescale. Ascending the giant branch, the starspot filling fraction drops dramatically as expected and is consistent with zero for most stars. However, we find a few interesting stars which we find to be highly spotted and radial velocity variable in the sub-subgiant population---we describe in detail the solution of one of these members, S1063, in Section \ref{sec:analysiss1063}.

There has been significant work to study the solar-type stars in M67 using activity proxies. \citet{2006ApJ...651..444G} measured Ca II H+K core strengths as an HK index and found that the majority ($72$–$80\%$) of solar analogues have values in the solar range with $17\%$ below and $7$–$12\%$ above, and variability over timescales longer than $6$ years. \citet{2018ApJ...855L..22B} used these data as evidence for anti-solar differential rotation, since models these models support larger absolute differential rotation.
Due to a sparsity in sources in common between APOGEE DR16 and literature chromospheric measurements in M67, there is insufficient data to mark a correlation for the solar-type stars; in our upcoming full catalog, we will have enough field star measurements to study activity in the solar regime.

\subsubsection{Starspots on an interacting sub-subgiant: S1063}\label{sec:analysiss1063}

Although we focus on single dwarfs in this work, it is also straightforward to interpret starspot filling fractions for binary stars with a minimal level of flux contribution from their companions. In M67, there are a number of heavily-spotted sub-subgiants, consisting of a binary system with an evolved primary \citep{2017ApJ...840...67L,2017ApJ...840...66G,2017ApJ...842....1G}. These are star cluster members with an anomalous position to the red of, and below, the evolved star cluster sequence. S1063 and S1113 have been identified by \citet{2017ApJ...840...66G} as sub-subgiants; they were also observed by APOGEE as part of the targeting strategy in M67. \citet{2022ApJ...925....5G} studied this star in detail with an observation with multiple IGRINS echelle orders, mapping this to four years of archival light curves from K2 and ASAS-SN. They observed that the starspot filling fraction was consistent with $32\%\pm7\%$ across nine spectral orders, with a spot temperature of $4000 \pm 200$ K and ambient temperature of $5000$–$5300$ K. With light curves from K2 and ASAS-SN, they model the cyclic variation in the starspot filling fraction as between $20$–$45\%$.

For S1063, our starspot estimate is $f_{\mathrm{spot}} = 20.3 \pm 0.4\%$. This is consistent with the lower end of the cyclic variation estimated from \citet{2022ApJ...925....5G} and may be the result of our analysis of just one APOGEE visit per star in this work. A systematically higher starspot estimate using the full IGRINS echelle spectra is also seen in Section \ref{sec:analysislkca4}---for LkCa 4 the echelle estimate is systematically higher than our H-band estimate, while individual echelle orders appear to agree with our measurement; a larger sample in common is needed to study whether such a systematic exists between the methods. We also estimate a value of $f_{\mathrm{spot}} = 35.8 \pm 1.1\%$ for S1113, which is significantly redder than S1063. New large samples of RS CVn's and sub-subgiants may enable new tests of stellar physics with interacting systems \citep[][Patton et al. in preparation]{2022ApJ...927..222L}.

\subsection{Accuracy and Precision of Two-Temperature Fits}\label{sec:accuracyprecision}

In the Pleiades we can attempt to characterize the random error in the technique in a young cluster by estimating the dispersion of measured starspot filling fractions in the saturated domain. From Figure \ref{fig:SP2aI_ple_rossby_final_tp_big} we see that this dispersion calculated for the saturated domain is $\sigma_{f_{\mathrm{spot}}} = 0.066$. It is uncertain whether this measured dispersion represents variability cycles in these stars or whether it is an uncertainty which is to be associated with the starspot estimation technique; however, as described in Section \ref{sec:spotrossby}, the fact that these stars are better described by a uniform than a normal distribution suggests that actual cycle variability is being measured. We note that a fiducial random error for our starspot measurements cannot be higher than this measured dispersion, and is likely lower. The measurement of $24.8\%$ as the fiducial starspot filling fraction at saturation for Pleiads also appears to be consistent with estimates from spectral indices in TiO \citep{2016MNRAS.463.2494F}.

In M67 most main sequence stars in our sample are solar-like and a measurement of $f_{\mathrm{spot}} = 3.0 \pm 4.4\%$ includes the entire population of main-sequence stars in Figure \ref{fig:SP2aI_CMDs_big}. For the solar-like stars, defined as a cut between 5700---5900 K, there are just ten stars with an average of $f_{\mathrm{spot}} \sim 2.0 \pm 1.1\%$. There also appears to be a clustering of points at or below $0.5\%$ spot filling fraction near solar age on the main sequence, close to the $0.3\%$ expected for solar maximum. On the other hand the detection of $0.3\% \pm 0.8\%$ starspot filling fraction for the giants strongly suggests that our technique is very precise in that domain; for these likely inactive giants, the flux contribution from a main sequence companion is miniscule, and a non-detection of spots is as expected. With the spectroscopic RV flags from \citet{2021AJ....161..190G}, these are likely almost all single stars, but as there are additional concerns in the higher mass stars at $T_{\mathrm{eff}} \gtrsim 6000$ K (such as gravity darkening, starspot distribution, or strong differential rotation), we caution against the interpretation of their spotted signal for now. However, the clear starspot signal in the subgiants at lower temperatures indicates a revival of activity at late times, which may be driven by evolutionary changes in the overturn timescale and internal angular momentum transport \citep{1985ApJ...299..286G}.

Despite the significantly lower starspot filling factor and dispersion in the solar-like stars, an estimate of the precision of this technique in inactive main sequence stars requires further analysis. We are developing additional grids tailored to inactive stars to characterize the starspot solution in these older main sequence stars.

\subsection{Stellar Parameter Systematics}\label{sec:systematics}

The seven-dimensional fit as described in Section \ref{sec:gridconstruction} is a joint fit to existing stellar parameters as well, including $T_{\mathrm{eff}}$, logg, $[M/H]$, $v \sin \, i$, and microturbulence. Identifying systematics between our parameters and existing measurements is necessary to characterize the two-temperature starspot solution.

\subsubsection{Systematics in Effective Temperature}\label{sec:systeff}

\begin{figure}
\includegraphics[trim={0cm 0cm 0cm 0cm},width=\columnwidth]{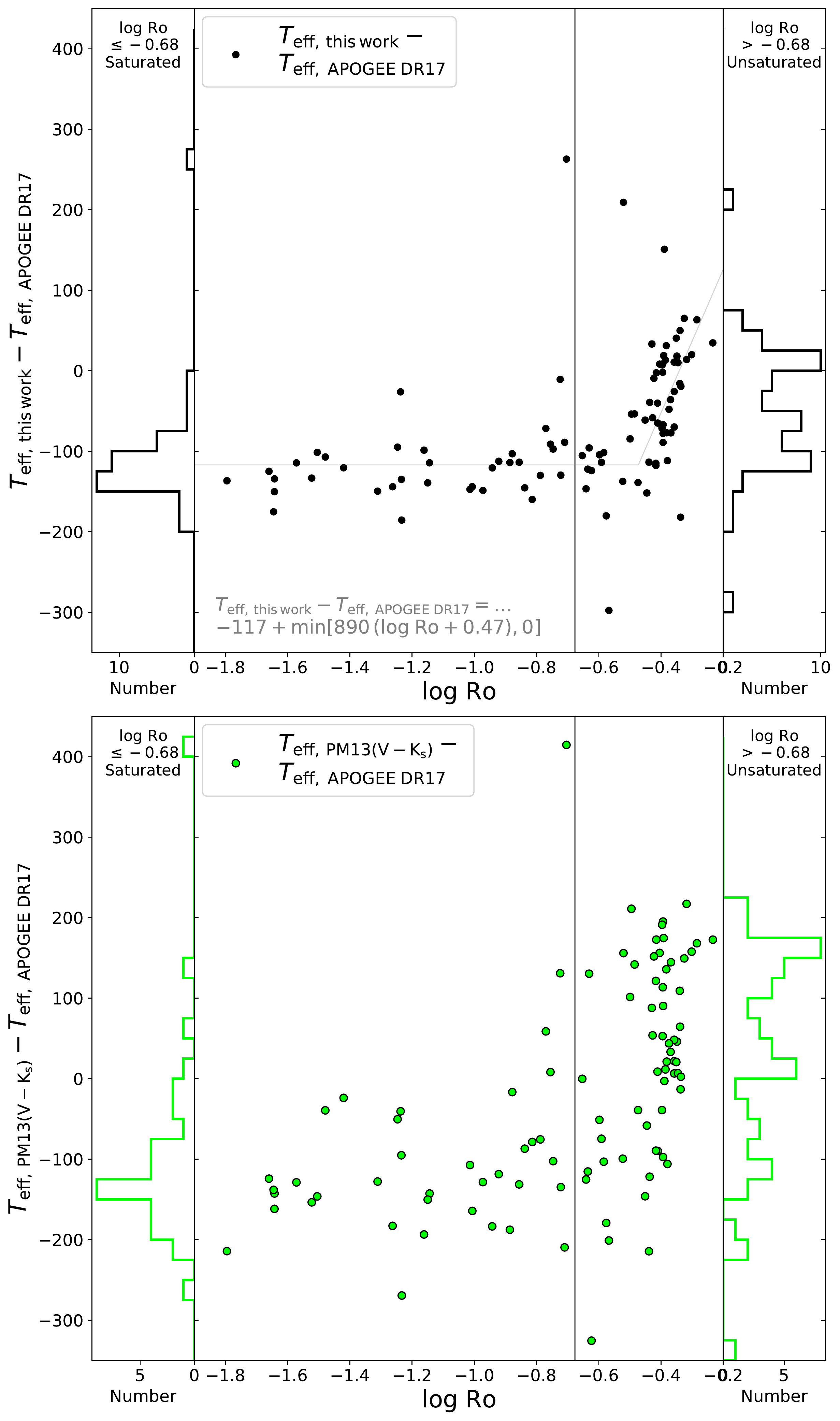}
\vspace{-1.5em}
\caption{\label{fig:SP2aI_sys_delTeff_big_pwr} Plots of the residual in $T_{\mathrm{eff}}$ against Rossby number between $T_{\mathrm{eff}}$'s derived in this work and from the \citet{2013ApJS..208....9P} system against APOGEE DR17. Top: $T_{\mathrm{eff}}$'s from this work against APOGEE DR17. Bottom: effective temperatures from $V-K_s$ estimated from the \citet{2013ApJS..208....9P} online extended dwarf table against APOGEE DR17. Left and right: histograms for the saturated and unsaturated stars, respectively.}
\end{figure}

In this work, we obtain two distinct temperature measurements using a two-temperature fit to spectra (Equation \ref{eqn:fluxtwoteff}); assuming flux conservation, the two components are related to the effective temperature through Equation \ref{eqn:twoteff}. Since the measured $T_{\mathrm{eff}}$ is a synthesis of two temperature components, these effective temperature solutions can differ fundamentally from a one-temperature model---since line fluxes can change between one- and two-temperature models. To demonstrate this, we compare our measurements to the crossmatched spectroscopic $T_{\mathrm{eff}}$ measurements from APOGEE DR17 \citep{2022ApJS..259...35A}. We also include an estimate of $T_{\mathrm{eff}}$ with a color-based method using the intrinsic colors of \citet{2013ApJS..208....9P}; we apply the same extended dwarf table as used in Section \ref{sec:spotproxy} on the dereddened $V-K_s$ color from \citet{2016AJ....152..113R} to infer this effective temperature. The difference between $T_{\mathrm{eff}}$ measurements from our work and APOGEE DR17 is shown as the top plot in Figure \ref{fig:SP2aI_sys_delTeff_big_pwr}; the bottom plot indicates the difference in $T_{\mathrm{eff}}$'s between a color-based $T_{\mathrm{eff}}$ from \citet{2013ApJS..208....9P} and APOGEE DR17 instead. We also include a maximum likelihood fit to the temperature residual---Rossby relationship as the gray model line and equation in the top panel of Figure \ref{fig:SP2aI_sys_delTeff_big_pwr}.

The systematic appears markedly different between the Rossby-saturated stars and unsaturated stars. The saturated stars show a strong negative bias in their effective temperatures relative to the system in APOGEE DR17, which is seen in both the effective temperatures estimated in this work and those from the \citet{2013ApJS..208....9P} system. The temperatures estimated from our two-temperature method and from the \citet{2013ApJS..208....9P} method are cooler than a single-temperature fit by $\sim$$100$ K. Quantitatively, this is $-109 \pm 11$ K for our measurements and $-98 \pm 19$ K for the color-based measurements; however, excluding the single outlier at high $\sigma$ in the color-based method with a $\Delta T_{\mathrm{eff}} > 400$ K results in a mean value of $-113 \pm 13$ K. In the unsaturated domain the offset between our measurements and that of APOGEE DR17 is at $-46 \pm 11$ K; the comparable offset for the unsaturated domain in the \citet{2013ApJS..208....9P} system is $33 \pm 17$ K instead. This shows that our effective temperatures are on the same system as \citet{2013ApJS..208....9P} for active stars.
The unsaturated stars also show a trend in temperature residuals with an activity dependence similar to the starspot---Rossby relation. Even moderate starspot coverage at the level of ($\sim$10\%, corresponding to $\log \mathrm{Ro} = -0.42$) imposes systematics on spectroscopic $T_{\mathrm{eff}}$'s of 70 K relative to the unspotted solution---suggesting that for stellar parameter estimation, a starspot treatment may be necessary for stars that have covering fractions near $10\%$ to ensure biases stay below the 70 K level. Additionally, the saturation threshold of the temperature residual---Rossby relation is at $\log \mathrm{Ro} = -0.47$, which from Figure \ref{fig:SP2aI_ple_mcmc_logRo_loc_ps_big} is within 1$\sigma$ from the Ca IRT and amplitude thresholds, but $>$2$\sigma$ discrepant from all other thresholds. As this temperature residual threshold is significantly higher than the Rossby threshold in starspots, above $\sim$$13\%$ spot coverage, even stars expected to be unsaturated in starspots may reach the maximal offset of 100 K in their effective temperatures.

The pattern in the effective temperature residuals indicates that two temperature components in the infrared spectra of active stars are effective at estimating global stellar parameters in concordance with the \citet{2013ApJS..208....9P} scale. As these stars are heavily spotted ($f_{\mathrm{spot}} \sim 25\%$), a single-temperature fit may consistently overestimate the temperature if it fits to the higher temperature component. Instead, a flux-conserving two-temperature method takes into account the cooler component (Equation \ref{eqn:twoteff}). We note that there may also be alternative reasons for why there may be a temperature offset, such as the PHOENIX model atmospheres used in this work compared to MARCS in ASPCAP for APOGEE DR17; however, this bias trends strongly with activity and starspot filling fraction on the host star, approaching zero for the unsaturated stars for which starspots become less important. This favors the explanation involving the difference in methodology which becomes more important in the active stars---the two-temperature fit this work and the single-temperature fit from ASPCAP.

\subsubsection{Systematics in Surface Gravity}\label{sec:syslogg}

Since our two-temperature method assumes that the two temperature components originate from the same star, the surface gravity used in the synthesis of each component is the same. As the logg's used in the synthesis do not change as the result of our starspot treatment, we do not expect large temperature-dependent systematics as the result of activity.

\begin{figure}
\includegraphics[trim={0cm 0cm 0cm 0cm},width=\columnwidth]{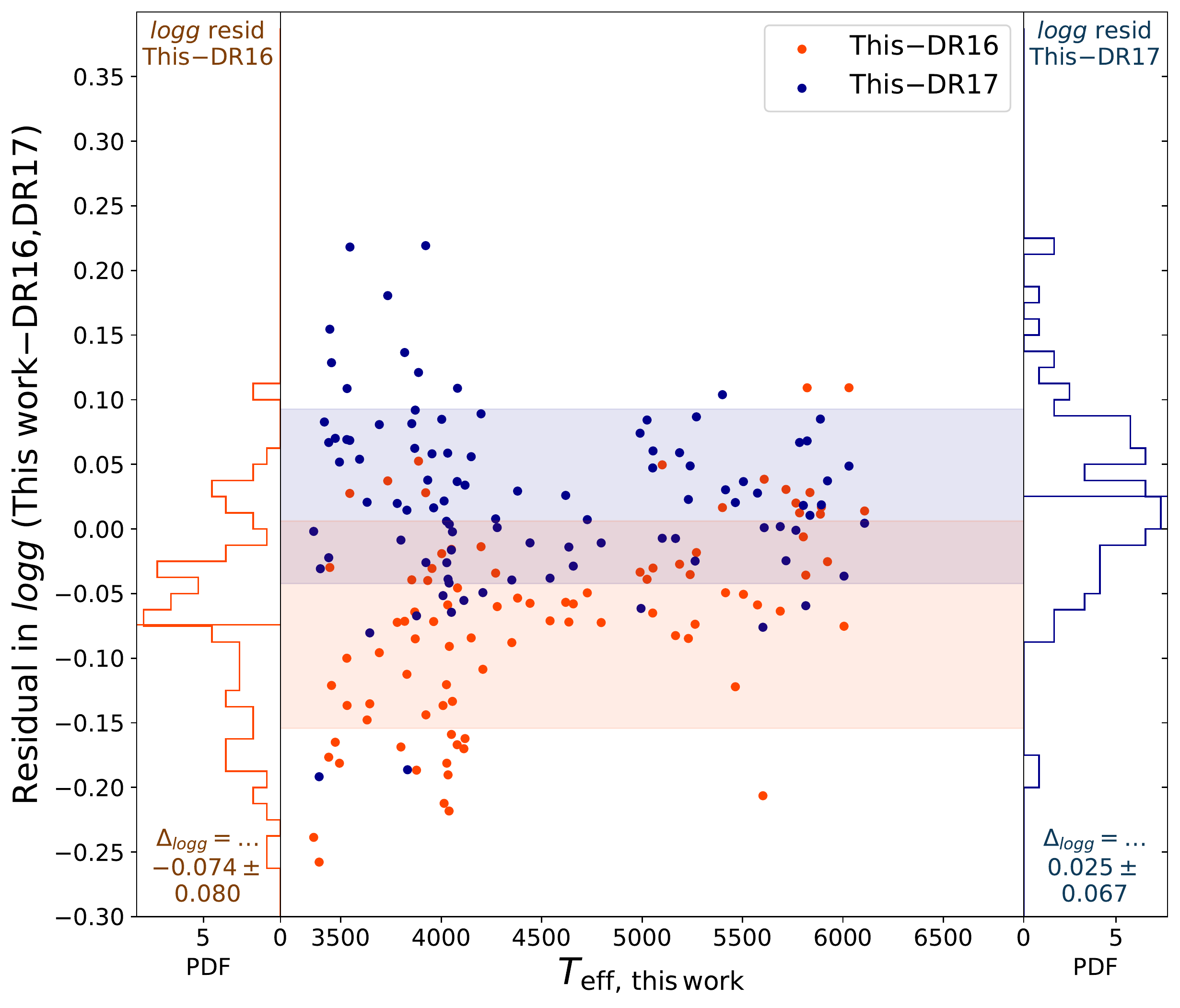}
\vspace{-1.5em}
\caption{\label{fig:SP2aI_sys_dellogg_big} The residual in logg between our calculation and the APOGEE DR16 (orange) and DR17 (blue) logg values. Left: histogram indicating the residual against DR16, including the mean and dispersion; right: histogram indicating the residual against DR17, including the mean and dispersion.}
\end{figure}

The surface gravity measurement from the pipeline is plotted in Figure \ref{fig:SP2aI_sys_dellogg_big} as a difference between our estimates and APOGEE DR16 (orange) or APOGEE DR17 (blue). The trend appears to be mass-dependent, with a larger offset for cool stars and a smaller offset for hot stars; our answers are situated between the DR16 and DR17 surface gravity measurements, albeit significantly closer to the DR17 scale ($0.025 \pm 0.007$ dex for DR17 compared to $-0.074 \pm 0.008$ dex for DR16). This similarity suggests that our logg measurements are consistent with similar inference methods using APOGEE near-IR H-band spectra. The increased scatter between the scales at the cool end represents the larger ($\sim$$0.2$ dex) systematic between cool star surface gravities in the Pleiades between DR16 and DR17.

The APOGEE DR17 surface gravity calibration was applied smoothly across evolutionary states from a neural network, whereas previous data releases were calibrated separately \citep{2022ApJS..259...35A}. We do not perform any kind of calibration on the raw logg output from our grid search in {\sc{ferre}}, so it is encouraging that we are in between the two calibrated surface gravity scales.

\subsubsection{Systematics in Metallicity}\label{sec:sysmh}

\begin{figure}
\includegraphics[trim={0cm 0cm 0cm 0cm},width=\columnwidth]{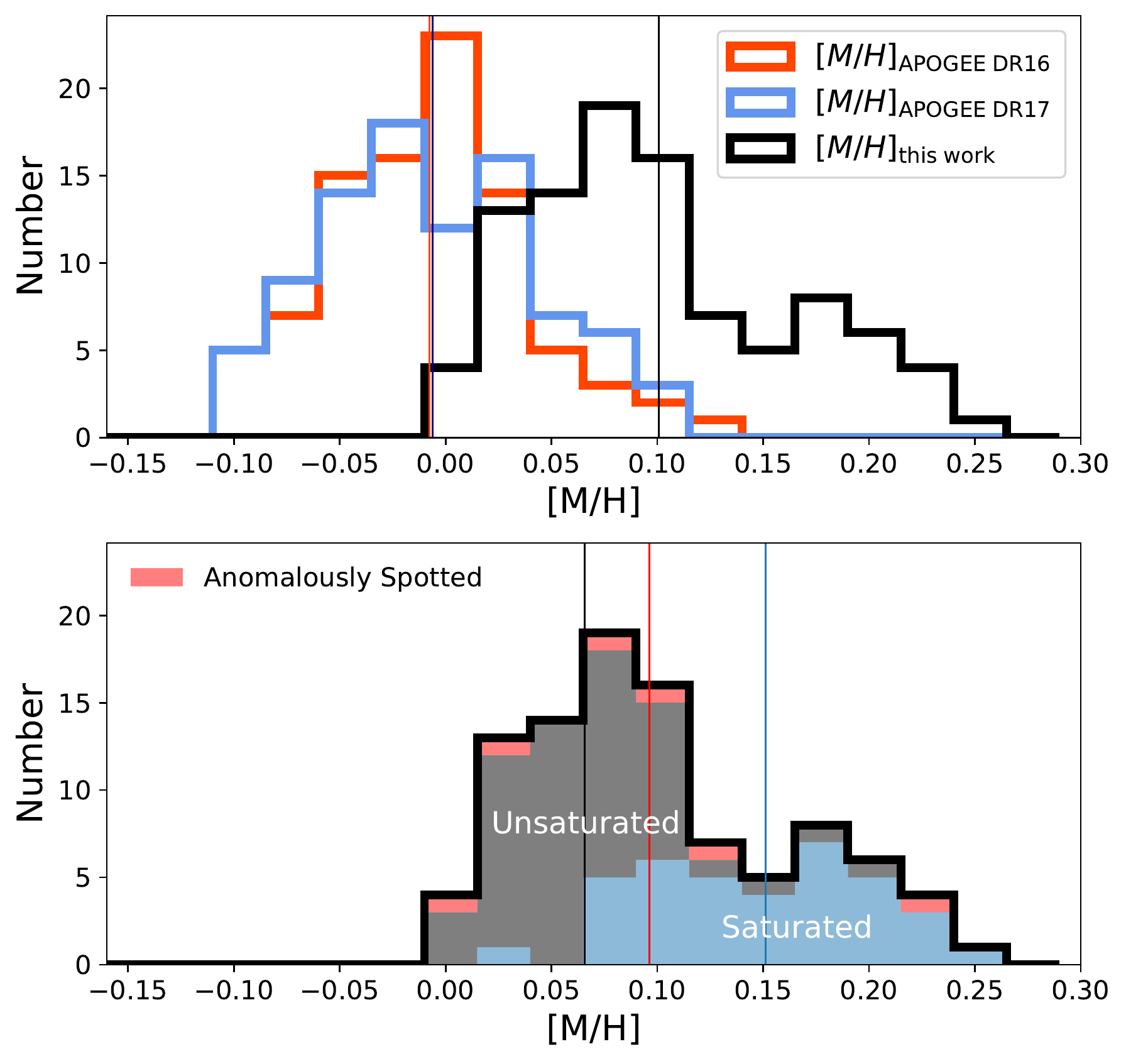}
\vspace{-1.5em}
\caption{\label{fig:SP2aI_sys_delMH_big} The distribution of Pleiades metallicities. Top: histograms corresponding to the distribution of Pleiades metallicities for APOGEE DR16 (red), DR17 (blue), and this work (black). Bottom: breakdown of metallicities into stacked histograms corresponding to saturated (blue), unsaturated (black), and anomalously spotted (red) stars. Means lines are overplotted with the corresponding color.}
\end{figure}

The distribution of metallicities for stars in the Pleiades is presented in Figure \ref{fig:SP2aI_sys_delMH_big}. The top panel shows three outlined distributions corresponding to the metallicity recovered from this work, APOGEE DR17, and APOGEE DR16. The bottom panel shows just the distribution of metallicities presented in this work, colored by activity state as identified previously.

From the top panel, we can see that the APOGEE metallicities are distributed similarly, and the metallicity estimates from this work are systematically higher. These distribution means are $-0.008 \pm 0.005$ dex in DR16, $-0.006 \pm 0.005$ dex in DR17, and $0.101 \pm 0.007$ dex in this work. As noted by \citet{2009AJ....138.1292S}, the metallicity of the Pleiades appears to be $\sim$0.03$\pm$0.02 with a $\pm$0.05 systematic error, with seven literature estimates ranging from slightly sub-solar to $0.06$ dex. At $0.10$ dex our measurements are higher than literature estimates, but are statistically consistent---considering systematic errors---at the $1\sigma$ level.

Breaking these populations into saturated, unsaturated, and anomalously spotted populations, it becomes clear that the unsaturated and saturated distributions are systematically offset from each other, at the level of $0.08$ dex. The corresponding means and standard errors are $0.066 \pm 0.006$ dex for Rossby-unsaturated stars, $0.151 \pm 0.010$ dex for the saturated stars, and $0.096 \pm 0.030$ for the anomalously spotted stars. The measurement for the Rossby-unsaturated stars appears to be similar to literature estimates \citep{2009AJ....138.1292S}, and the systematic for the saturated stars appears to come with a larger random error as well. We note however that we are seeing a similar systematic in metallicity as seen by \citet{2022ApJ...927..123S}, who found a $\Delta$ [Fe/H] of $0.10 \pm 0.04$ dex for the same spectra between ASPCAP parameters and their work using the \citet{2021AJ....161..254S} linelist. As this systematic in the metallicities appears to be small, and appears to be consistent with previous studies, we conclude that this systematic is not likely due to any methodological changes.

\subsubsection{Systematics in Rotational Velocity}\label{sec:sysvsini}

\begin{figure}
\includegraphics[trim={0cm 0cm 0cm 0cm},width=\columnwidth]{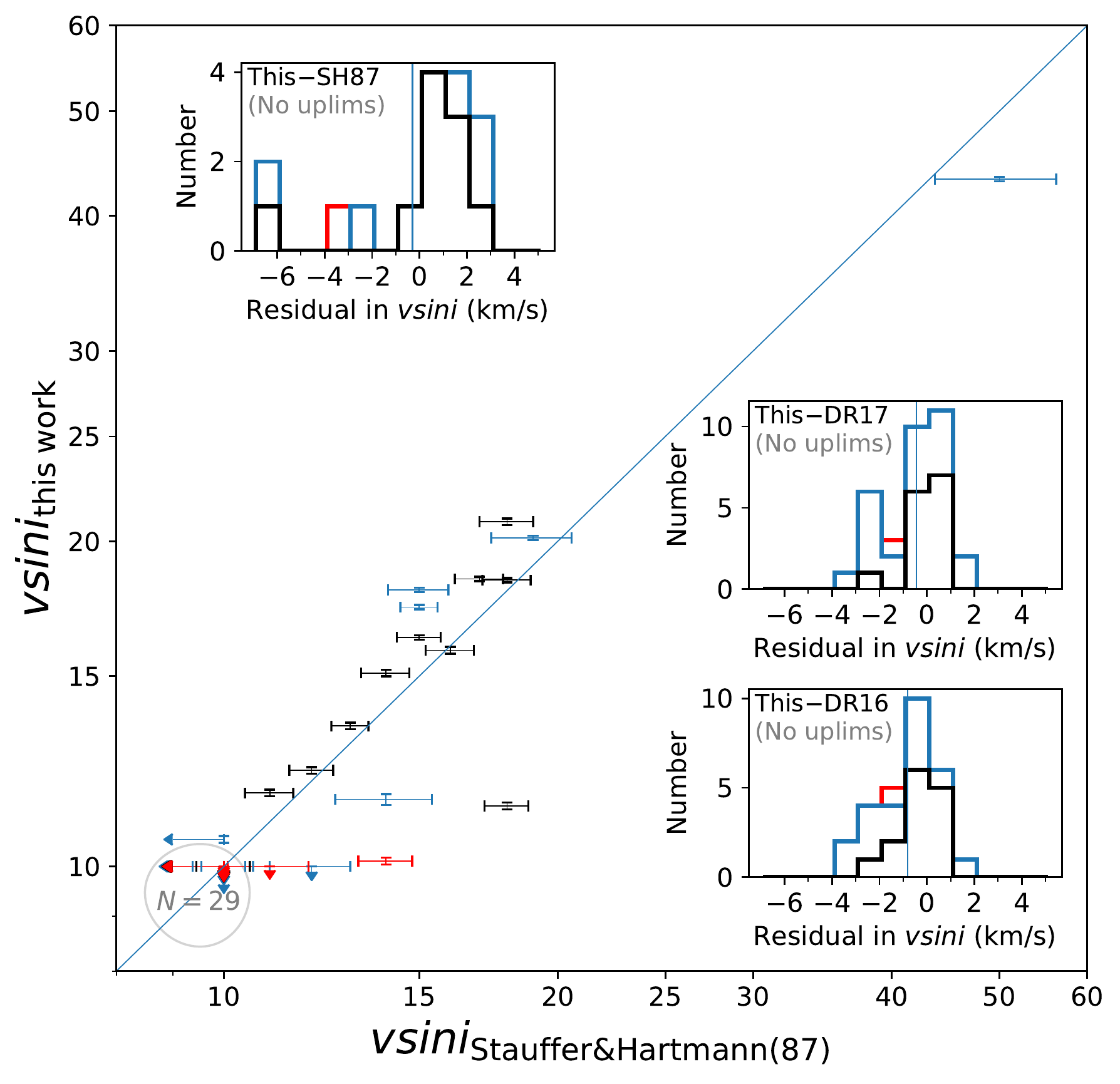}
\vspace{-1.5em}
\caption{\label{fig:SP2aI_sys_delvsini_big} Correspondence between estimated $v \sin \, i$ from \citet{1987ApJ...318..337S} (x-axis) and our estimates of $v \sin \, i$ (y-axis). Inset plots show the histograms of $v \sin \, i$ residuals between our measurements and \citet{1987ApJ...318..337S} (top left), our measurements and APOGEE DR17 (middle right), and our measurements and APOGEE DR16 (bottom right).}
\end{figure}

Finally, we demonstrate the offset in $v \sin \, i$ between our recovered values and those reported from \citet{1987ApJ...318..337S} in Figure \ref{fig:SP2aI_sys_delvsini_big}, color-coded by Rossby saturation regime. The inset histograms show the offset in $v \sin \, i$ between our measurements and the \citet{1987ApJ...318..337S}, APOGEE DR17, and APOGEE DR16 reported measurements. We exclude any $v \sin \, i$ measurements with upper limits in these residual histograms as including them may artificially bias the distribution low, and we adopt an uncertainty on the \citet{1987ApJ...318..337S} $v \sin \, i$ measurements of $v \sin \, i / 2(1+R)$ as suggested, where $R$ is their ratio between their cross-correlation peak and antisymmetric component. Between all four systems, the $v \sin \, i$ measurements appear to be generally consistent to better than 1 km/s. The residuals and standard errors are $-0.30 \pm 0.74$ km/s between our measurements and \citet{1987ApJ...318..337S}, $-0.37 \pm 0.18$ km/s between this work and DR17, and $-0.66 \pm 0.20$ km/s between this work and DR16.

With all offsets in measured rotational velocity being within $1$ km/s of published catalogs, we conclude that our $v \sin \, i$ measurements are consistent with the literature. In terms of the reported rotational velocity errors, the mean random error for our Pleiades single star sample is $0.15$ km/s, and the mean systematic between this work and the DR16, DR17, and \citet{1987ApJ...318..337S} measurements is approximately $-0.44$ km/s. This consistency shows that the starspot solution does not appreciably affect the measurement of rotational velocities from spectra in single stars.

\section{Discussion}\label{sec:discussion}

The detection of a saturation in inferred starspot filling fraction and an associated decline at higher Rossby number in Figure \ref{fig:SP2aI_ple_rossby_final_tp_big} is a strong indication that starspots saturate for the most rapid rotators. It has been seen throughout the literature that chromospheric, coronal, and magnetic measurements saturate at rapid rotation, but it has remained an open question whether the saturated chromosphere and ordered magnetic fields produce a fixed maximum value in starspot coverage as seen through spectroscopy. With this work, we conclusively detect a maximum starspot filling fraction which is attained at a similar rotation rate as seen in other indicators. Our equipartition magnetic field estimates also see a saturation at rapid rotation, which has also been recently verified in M dwarfs through the Zeeman broadening technique \citep{2022A&A...662A..41R}. In this work we also show that the starspot parameter is well-correlated with other chromospheric and coronal activity proxies.

However, these starspot measurements have newly made it possible to probe the structure of stellar surface dynamos in detail. We indicate that starspots and magnetism affect stellar structure and change the convective turnover timescale in a systematic manner \citep{2020ApJ...891...29S}, an effect that we model consistently with these spectroscopic starspot measurements. We have shown in this work that the effect on Rossby numbers is maximized for active stars, systematically shifting the Rossby threshold at saturation. Rather than being used as an empirical fit parameter, we argue that changes in convective turnover timescale predicted by stellar evolution theory must be accounted for to test the Rossby paradigm as stars evolve. The over-spotted single Pleiads identified in this work challenge the traditional Rossby consensus that all slow rotators are inactive---since the gap between observations is on the order of decades and likely uncorrelated, we note that it is unlikely that the detected over-spotted stars are simply varying due to activity cycles. Placing these over-spotted stars in a framework of stellar evolution may challenge the uniformity of the activity---Rossby relation, and lead to more sensitive diagnostics of stellar physics.

The interplay between starspot filling fraction and other activity proxies can now be explored quantitatively to test stellar dynamo theories; testing whether the changing saturation threshold of different activity proxies is the result of phenomena occurring at different convective depths or intrinsically predicted by stellar theory can now be explored for large populations. In particular, the comparison between our starspot filling fractions and equipartition magnetic field strengths can now be explored as a fundamental prediction of stellar evolutionary models. The comparison between these measurements and ZDI or Zeeman broadening estimates, which are sensitive to the large-scale signed and small-scale unsigned fields respectively, may test scenarios of equipartition or spot geometry.

The question of whether cycle variation can describe stars which scatter off the mean activity---Rossby relation can now be tested; in this work, we found using a K-S test that the distribution of starspot filling fractions for the active Pleiads was too broad to be fit by a sum of two Gaussians. Instead, a uniform distribution with a smaller random error component more appropriately describes the distribution, with a p-value significantly above $0.05$. This appears to suggest that our starspot measurement is more precise than the population dispersion (which might otherwise impose a Gaussian error on the distribution) and that we are seeing some degree of cycle variation. With spectroscopic starspot filling fractions for large populations of stars and repeat visit spectra, it will be possible to investigate the range of starspot variation across activity cycles, and whether certain classes of stars are shifted relative to the mean relation.

We also find that this technique is robust; it is suitable not only the activity of active young stars, but also recovers starspot filling fractions for stars of different evolutionary states in M67. The solar-like starspot filling fractions for the solar analogues, the non-detections in the inactive old giants, the rising activity in subgiants, and the high spot coverages in the interacting sub-subgiants suggests that this tool is broadly applicable in testing stellar physics across different evolutionary states. For instance, tracking solar variability with large samples of solar analogues in APOGEE, detecting magnetism on giants, testing theories lengthening the convective turnover timescale in subgiants, and probing physics of interacting systems are all examples of applications of this method. We validate the prediction across a large sample that lengthening convective turnover timescales in subgiants leads to an increase in activity from the main sequence \citep{1985ApJ...299..286G}, and we find concordance between our measurements and a number of existing starspot measurements in the literature in the young star LkCa 4, the active Pleiad HII 296, and the interacting sub-subgiant S1063.

This work jointly fits for global stellar parameters, and we explore the offsets in effective temperature, surface gravity, metallicity, and rotational velocity against reported literature values. The effective temperatures reported by our method are consistent with the spectroscopic ASPCAP solution in APOGEE DR17 for the inactive stars, but are cooler by $\sim$$100$ K for the active stars. A detailed analysis finds that the temperature residuals saturate and decline similar to other activity---Rossby relations, with a correlated increase of the absolute value of the temperature residual with activity; even in Rossby-unsaturated stars, a level of $10\%$ spot coverage is enough to result in a 70 K systematic. For a star at $4000$ K this temperature systematic results in an inflation of the inferred radius by $3.5\%$---suggesting that inflated radii in active stars is a generic result from this technique. We find that this temperature offset places our effective temperature measurements on the \citet{2013ApJS..208....9P} temperature scale, meaning that our spectroscopic $T_{\mathrm{eff}}$'s are now consistent with temperatures derived from SED fitting over a curated sample of stars. This is likely the result of a two-temperature fit also identifying the spot component in the most spotted stars, which may have been previously fit inadequately with a single temperature in spectra. The surface gravities reported in this work are intermediate between the ASPCAP solutions in APOGEE DR16 and DR17, appearing to be closer to the DR17 system even without any calibration procedure. We find that our metallicities are on the high end but still consistent with literature estimates for the Pleiades, particularly in the inactive stars; however, we find a systematic in the metallicity estimates in the active stars of $0.08$ dex. A systematic of similar magnitude was also seen in the M dwarfs in \citet{2022ApJ...927..123S}. Because our saturated stars are later in spectral type than the less active ones, this may not be an activity bias, but could instead be tied to issues with the linelists for cool stars. Finally, we find that our rotational velocity measurements are consistent with \citet{1987ApJ...318..337S} and the ASPCAP solutions in APOGEE DR16 \& 17 with an estimated systematic error of $-0.44$ km/s.

Our method has limitations---stars have complicated surface structures including starspots, network, and faculae; the empirical APOGEE linelists are still not complete, and the starspot estimates in H-band spectra may have systematics compared to other spectral ranges. Starspots change across a cycle and over time, which means that a single snapshot may not fully describe the activity cycle for non-simultaneous measurements. The choice of model atmosphere and the lack of limb-darkening treatment in this work may impose systematics, though they are likely small. We may not be measuring starspots in higher mass stars, where gravity darkening may be an issue. Binary contamination is also an important issue to resolve for these two-temperature measurements, which we mitigate using kinematics from {\gaia} and APOGEE as well as photometric binary rejection. However, even with these limitations of this approach, we detect strong starspot and magnetic signals which appear to be very precise and accurate in open clusters; we also jointly provide stellar parameters which appear to be consistent with prior work.

\section{Conclusions}\label{sec:conclusions}
Our main result is that we have developed a practical tool, applicable to a large survey dataset, for measuring starspot filling fractions and estimating equipartition magnetic field strengths in cool stars. We recover temperature contrasts consistent with starspots even without imposing them by assumption; we get starspot---Rossby and $\left< \left| B_{\mathrm{eqp}} \right| \right>$---Rossby patterns similar to those from traditional activity measures; and we recover low filling fractions in stars that are not expected to be active.

With these measurements, it is possible to probe the morphology and evolution of the activity---Rossby relation. In this work, we have shown that there are stars which appear anomalously spotted and high in chromospheric and coronal activity proxies, which are not obviously binaries nor cycle-variable stars seen in a high state. With a large set of single-star measurements, we will enable tests of whether the activity---Rossby plane is a function solely of Rossby number or whether there are additional degrees of freedom for activity state. These data will help uncover whether fully convective stars, synchronized binaries, solar-like stars, and evolved stars have activity signatures which behave similarly at different Rossby numbers. We will also be able to test whether predictions of stellar theory, such as core-envelope decoupling, have observational signatures. The evolution of the starspot---Rossby relation with a large survey dataset may uncover new physics at a host of different ages, and have applications not only to stellar physics but also to exoplanet habitability.

Presenting a survey-sized catalog of not only starspot filling fractions but also magnetic field estimates for stars will also help guide simulations of stellar surfaces. The comparison between our magnetic detections and those inferred from complementary Zeeman measurements may prove to be useful in probing the geometry of stellar magnetic fields, or the systematics inherent in each technique. Multiple APOGEE visits may also probe starspot variability across the time domain, and study whether starspot and magnetic cycle variation changes in active or inactive stars. With a large range of solar-type stars and a more sensitive starspot grid, we may also be able to test the variation of the solar cycle in the infrared. With our jointly estimated $v \sin \, i$, the polar or equatorial spot distribution may even be inferred. The method presented in this paper will permit quantitative studies of these and more features in stellar dynamo theory across large populations of stars in APOGEE.

The accurate spectroscopic stellar parameters inferred from this work account for the structural impact of starspots on the star. This has previously been shown to have a strong effect on the radii of active stars, and for the ages and masses of young stars. We present $T_{\mathrm{eff}}$'s which are newly consistent with the \citet{2013ApJS..208....9P} temperature scale for all stars, most notably the active stars, which previously showed a systematic bias in single-temperature spectroscopic measurements. This also has applications for fields which study active stars and derived stellar parameters; a homogeneous set of effective temperatures can be quite useful for calibrating stellar evolution theory for all active stars and in modelling exoplanetary parameters around active stars. This dataset is a complementary tool for galactic archaeology with APOGEE and can help test whether activity affects stellar parameters at different metallicity or elemental abundances.

We demonstrate throughout this paper a new set of accurate and precise starspot and equipartition magnetic field strength measurements utilizing APOGEE high-resolution spectra. These spectroscopic starspot measurements are robust and agree well with literature measurements of activity. The potential applications of this dataset are vast, with a set of starspot measurements and activity indicators which will soon be available for up to $\sim$650,000 stars over $\sim$2.6 million visits in APOGEE DR17, increasing the literature starspot and magnetic measurements by three orders of magnitude \citep{2022ApJS..259...35A}. Applications to datasets in Milky Way Mapper include $\gtrsim$6 million stars, including dedicated projects such as the solar neighborhood census for $\sim$400,000 starspot measurements for nearby $<100$ pc cool stars, $\sim$250,000 red giants, and $\sim$40,000 young stars \citep{2017arXiv171103234K}. With concurrent spectroscopy using the low resolution optical BOSS instrument, it will be newly possible to correlate activity proxies from strong emission lines to starspot measurements and probe stellar activity variations for large statistical samples. Additionally, it is possible that this technique can be applied in optical spectra as well, where it may not be sensitive to large regions of cool spots but instead to small regions of hot faculae. Thus this technique coupled with rotation datasets such as Kepler/K2, TESS, ZTF, and ASAS-SN is a powerful new tool in the study of stellar physics.

\section*{Acknowledgements}
The authors wish to thank Jennifer van Saders and Lee Hartmann for helpful comments on the manuscript in preparation. We thank Aline Vidotto and Ansgar Reiners for useful discussions. L.C. appreciates Jennifer Johnson and the broader SDSS collaboration for their continuing support and interest. The authors also thank Keivan Stassun for his insightful comments which have improved the manuscript in review. L.C. also wishes to thank the late Cao Fulin, whose inspiration made this work possible. The authors also acknowledge the computing resources and support made available by the Ohio State University College of Arts and Sciences. M.H.P. and L.C.
acknowledge support from NASA grant 80NSSC19K0597.

Funding for the Sloan Digital Sky 
Survey IV has been provided by the 
Alfred P. Sloan Foundation, the U.S. 
Department of Energy Office of 
Science, and the Participating 
Institutions. 

SDSS-IV acknowledges support and 
resources from the Center for High 
Performance Computing  at the 
University of Utah. The SDSS 
website is www.sdss.org.

SDSS-IV is managed by the 
Astrophysical Research Consortium 
for the Participating Institutions 
of the SDSS Collaboration including 
the Brazilian Participation Group, 
the Carnegie Institution for Science, 
Carnegie Mellon University, Center for 
Astrophysics | Harvard \& 
Smithsonian, the Chilean Participation 
Group, the French Participation Group, 
Instituto de Astrof\'isica de 
Canarias, The Johns Hopkins 
University, Kavli Institute for the 
Physics and Mathematics of the 
Universe (IPMU) / University of 
Tokyo, the Korean Participation Group, 
Lawrence Berkeley National Laboratory, 
Leibniz Institut f\"ur Astrophysik 
Potsdam (AIP),  Max-Planck-Institut 
f\"ur Astronomie (MPIA Heidelberg), 
Max-Planck-Institut f\"ur 
Astrophysik (MPA Garching), 
Max-Planck-Institut f\"ur 
Extraterrestrische Physik (MPE), 
National Astronomical Observatories of 
China, New Mexico State University, 
New York University, University of 
Notre Dame, Observat\'ario 
Nacional / MCTI, The Ohio State 
University, Pennsylvania State 
University, Shanghai 
Astronomical Observatory, United 
Kingdom Participation Group, 
Universidad Nacional Aut\'onoma 
de M\'exico, University of Arizona, 
University of Colorado Boulder, 
University of Oxford, University of 
Portsmouth, University of Utah, 
University of Virginia, University 
of Washington, University of 
Wisconsin, Vanderbilt University, 
and Yale University.

This work has made use of data from the European Space Agency (ESA) mission
{\gaia} (\url{https://www.cosmos.esa.int/gaia}), processed by the {\gaia}
Data Processing and Analysis Consortium (DPAC,
\url{https://www.cosmos.esa.int/web/gaia/dpac/consortium}). Funding for the DPAC
has been provided by national institutions, in particular the institutions
participating in the {\gaia} Multilateral Agreement.

This publication makes use of data products from the Two Micron All Sky Survey, which is a joint project of the University of Massachusetts and the Infrared Processing and Analysis Center/California Institute of Technology, funded by the National Aeronautics and Space Administration and the National Science Foundation.

\section*{Data Availability}
The original data and models underlying this article are available upon request. The full suite of models will be available publicly with the release of the catalog paper (Cao et al. in preparation).

\appendix
\section{Saturation Thresholds with Empirical Rossby Numbers}\label{sec:appendixrossby}
Throughout our analysis we use the self-consistent theoretical treatment of Rossby numbers described in Section \ref{sec:rossbystarspots}. However, alternative Rossby prescriptions are prevalent in the literature; one such definition is from \citet{2011ApJ...743...48W}. We would expect from Figure \ref{fig:SP2aI_Ro_Ro_big} that fits involving empirical Rossby numbers on the same scale may have saturation thresholds lower than our SPOTS Rossby numbers by $\sim$$0.1$–$0.2$ dex. We might wonder about the location of the flagged anomalous stars under an alternative definition of Rossby number. Do anomalous stars still appear anomalously to the top-right of the Rossby curve with an empirical Rossby number? And which of the theoretical or empirical prescriptions provides a cleaner sequence for purposes of fitting activity---Rossby relationships?

\begin{figure}
\includegraphics[trim={0cm 0cm 0cm 0cm},width=\columnwidth]{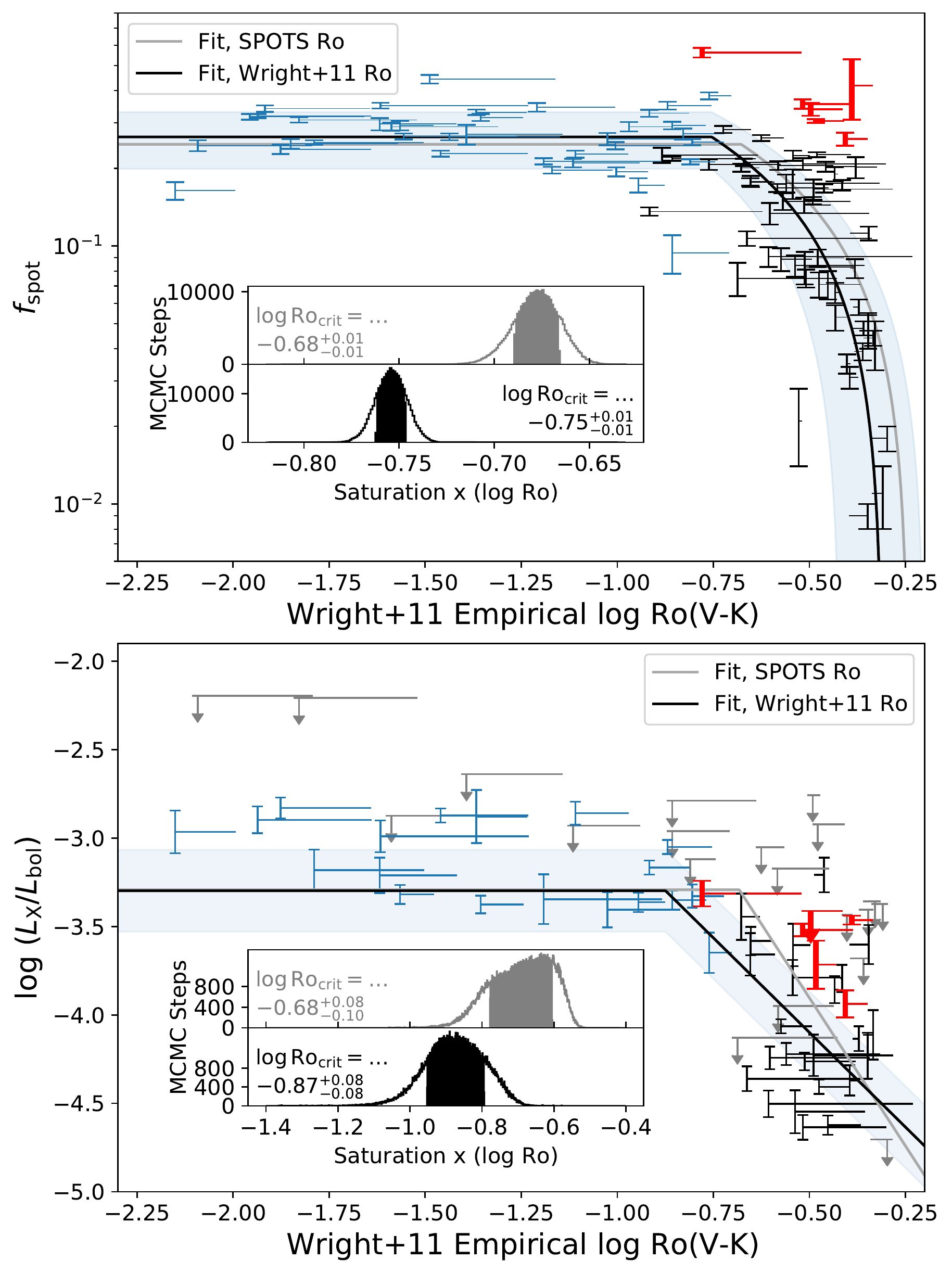}
\vspace{-1.5em}
\caption{\label{fig:SP2aI_ple_Ro_c_big} Effects of the alternative empirical Rossby numbers on our derived saturation thresholds. Top: fit comparison in starspot filling fraction. Bottom: fit comparison in X-rays. Lines indicate the offset in log Rossby number between the empirical Rossby number from \citet{2011ApJ...743...48W} and the theoretical one from \citet{2020ApJ...891...29S}; gray indicates the best fit relations from Sections \ref{sec:spotrossby} \& \ref{sec:rossbysat}, while black corresponds to the best fit solution to the empirical Rossby numbers as shown in these plots. Inset plots show the distribution of saturation thresholds corresponding to the theoretical Rossby number from SPOTS (gray) and empirical Rossby number (black).}
\end{figure}

We do direct comparisons on our sample, performing fits in $f_{\mathrm{spot}}$ and $L_{\mathrm{X}}/L_{\mathrm{bol}}$ using empirical Rossby numbers in Figure \ref{fig:SP2aI_ple_Ro_c_big}. The difference between each Rossby number from the V-K empirical relation and SPOTS is indicated by a horizontal line, showing offsets of $\sim$$0.2$ dex for the active stars and $\sim$$0$ for the inactive ones. In the top panel, the empirical Rossby definition yields a critical Rossby threshold offset in $f_{\mathrm{spot}}$ by $-0.07$ dex; in the bottom, the alternative definition results in an offset of $-0.20$ dex instead, in X-rays.

The systematic offset shown in this figure is the result of a treatment which accounts for the structural effect of starspots; current methods in the literature do not account for this effect, which can show up as a differential stretching of the activity---Rossby diagram for the most active stars, and an underestimate in spotted stellar Rossby numbers. Since unspotted and spotted methods show a trend in Rossby numbers and a bias in the Rossby number identified at saturation, we indicate that a proper transformation onto the same Rossby scale is required to compare derived activity---Rossby fit parameters.

\section{Chromospheric System Offsets}\label{sec:appendixchromospheric}
\begin{figure}
\includegraphics[trim={0cm 0cm 0cm 0cm},width=\columnwidth]{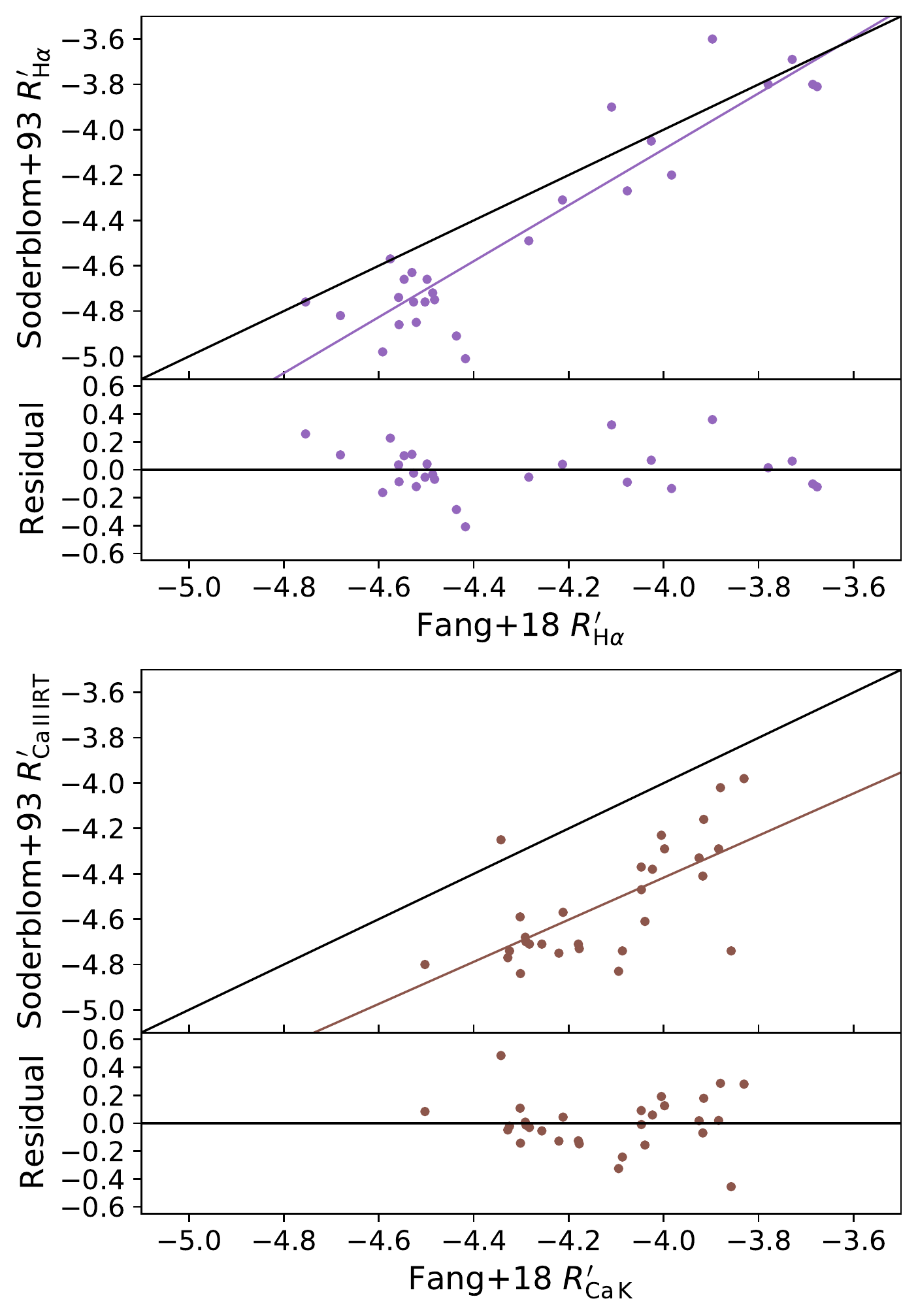}
\vspace{-1.5em}
\caption{\label{fig:SP2aI_offsets_td_big} Offsets between the photometric and spectroscopic activity relations of \citet{1993ApJS...85..315S} and \citet{2018MNRAS.476..908F} respectively. Top: a calibration from the LAMOST spectroscopic activity to the \citet{1993ApJS...85..315S} photometric system in the H$\alpha$ feature for stars in common in the Pleiades. Bottom: a calibration to the Ca II infrared triplet from the LAMOST Ca II K measurement \citep{2018MNRAS.476..908F} for stars in common in the Pleiades. A strong relationship between the Ca IRT and Ca H\&K lines is seen in the literature \citep{2017A&A...605A.113M}. Since they are different emission lines, we caution against the interpretation of and indicate clearly all plots where we use the transformed \citet{2018MNRAS.476..908F} calcium data.}
\end{figure}

For the H$\alpha$ measurements, we calibrate from the spectroscopic activity excess measurements provided by \citet{2018MNRAS.476..908F} to the scale established by the photometric $R^{\prime}_{\mathrm{H}\alpha}$ from \citet{1993ApJS...85..315S} using stars in common between the two samples in the Pleiades. While variation in activity due to activity cycles may be a concern for these stars, it is unlikely that the transformation locus is significantly perturbed by cyclic variation for the same stars. It is more likely that the calibration procedure reduces systematics between the two different techniques to infer $R^{\prime}_{\mathrm{H}\alpha}$.

We also do an empirical activity transformation from the spectroscopic inference of the parameter $R^{\prime}_{\mathrm{Ca \; K}}$ from \citet{2018MNRAS.476..908F} to the measurements of the \citet{1993ApJS...85..315S} infrared triplet $R^{\prime}_{\mathrm{IRT}}$. Transforming between Ca II H\&K and Ca II IRT is motivated by the tight correlation between the strong optical H\&K lines and the infrared triplet \citep{2017A&A...605A.113M}. As these are not the same line there may be issues with interpreting the excess after having transformed the LAMOST data. Because of this concern, we clearly note where these transformed data points are plotted and advise caution interpreting plots with Ca IRT transformed in this manner. We note that uniform Ca infrared triplet data is now available in LAMOST \citep{2021ApJS..253...51L} and is possible also with {\gaia} DR3 \citep{2022arXiv220605766L} as an activity index for large samples of nearby stars.

\bibliographystyle{mnras}
\bibliography{cite}

\bsp	
\label{lastpage}
\end{document}